# ACCRETION OF WATER IN CARBONACEOUS CHONDRITES: CURRENT EVIDENCE AND IMPLICATIONS FOR THE DELIVERY OF WATER TO EARLY EARTH

Josep M. Trigo-Rodríguez[1,2], Albert Rimola[3], Safoura Tanbakouei[1,3], Victoria Cabedo Soto[1,3], and Martin Lee[4]

[1] Institute of Space Sciences (CSIC), Campus UAB, Facultat de Ciències,
Torre C5-parell-2ª, 08193 Bellaterra, Barcelona, Catalonia, Spain. E-mail: trigo@ieec.uab.es

[2] Institut d'Estudis Espacials de Catalunya (IEEC), Edif.. Nexus,
c/Gran Capità, 2-4, 08034 Barcelona, Catalonia, Spain

[3] Departament de Química, Universitat Autònoma de Barcelona, 08193 Bellaterra, Catalonia, Spain.
E-mail: albert.rimola@uab.cat

[4] School of Geographical and Earth Sciences, University of Glasgow, Gregory Building, Lilybank Gardens, Glasgow G12 8QQ, UK.

**Abstract**:
Protoplanetary disks are dust-rich structures around young stars. The crystalline and amorphous materials contained within these disks are variably thermally processed and accreted to make bodies of a wide range of sizes and compositions, depending on the heliocentric distance of formation. The chondritic meteorites are fragments of relatively small and undifferentiated bodies, and the minerals that they contain carry chemical signatures providing information about the early environment available for planetesimal formation. A current hot topic of debate is the delivery of volatiles to terrestrial planets, understanding that they were built from planetesimals formed under far more reducing conditions than the primordial carbonaceous chondritic bodies. In this review, we describe significant evidence for the accretion of ices and hydrated minerals in the outer protoplanetary disk. In that distant region highly porous and fragile carbon and water-rich transitional asteroids formed, being the parent bodies of the carbonaceous chondrites (CCs). CCs are undifferentiated meteorites that never melted but experienced other physical processes including thermal and aqueous alteration. Recent evidence indicates that few of them have escaped significant alteration, retaining unique features that can be interpreted as evidence of wet accretion. Some examples of carbonaceous chondrite parent body aqueous alteration will be presented. Finally, atomistic interpretations of the first steps leading to water-mediated alteration during the accretion of CCs are provided and discussed. From these new insights into the water retained in CCs we can decipher the pathways of delivery of volatiles to the terrestrial planets.

# 1. Introduction

Protoplanetary disks are dust-rich structures around young stars. They are flattened clouds of billions of particles whose compositions vary in response to environmental gradients created by intense electromagnetic and particle irradiation from the young star (see e.g., Martínez-Jiménez et al., 2017). There is growing evidence that stars like the Sun are often formed in stellar associations or even clusters (Trigo-Rodríguez et al., 2009). The dust-rich environments surrounding young stars have been observed, and reviewed extensively (see e.g., Armitage, 2011) but information about the accretionary processes at work in the outer regions of these disks, where they pass into interstellar space, are still lacking. It is in such regions that the parent bodies of the carbonaceous chondrite meteorites (CCs) are believed to have formed (Weisberg et al., 2006). The lead-lead isotopic chronometer indicates that these rocks consolidated about 4.56 Ga ago from mineral condensates formed from the vapor phase at least since 4.567 Ga (Amelin et al., 2002). A current hot topic of debate is the delivery of water to terrestrial planets that were formed under far more reducing conditions than carbonaceous chondritic bodies (Alexander et al., 2018)

Disk instabilities, probably induced by the formation of giant planetary embryos, contributed to creating turbulence and size-sorting of disk materials (Boss, 2013). This process has been observed as noticeable dust clustering forming distinguishable rings in protoplanetary disks; *e.g.*, the disk recently observed in HL Tau by the Atacama Large Millimeter/submillimeter Array (ALMA) (Zhang et al., 2015). The objects formed are no larger than few tens of kilometers so that the heat released from their radioactive components is quickly irradiated to space. These bodies have escaped chemical segregation (i.e., are undifferentiated) and so retain primordial disk components.

Meteorite evidence suggests that planetesimals became kilometer-sized bodies in the inner disk within about 10 millions of years from the formation of Ca- and Al-rich inclusions (CAIs). These planetesimals accreted to make planetary embryos, from which the terrestrial planets formed over timescales of few tens of millions of years. This epoch was marked by huge impacts that compacted and sculpted asteroids, transforming many such objects into collisional shocked, brecciated, and complex rubble piles (Trigo-Rodríguez and Blum, 2009). Such processes, however, affected not all of them. Some asteroids stored in the outer regions of the disk were probably less highly collisionally processed, and are enriched in organic and volatile compounds, as revealed the study of the comets 81P/Wild 2 and 67/Churyumov-Gerasimenko (Brownlee et al., 2006; Schulz et al., 2015). Remarkably, laboratory simulations of materials accreting at similar relative velocities end with meteorite proxies that are highly porous and fluffier than current chondrites (see e.g., Blum et al., 2006).

As planetesimals were the building blocks of planets, it follows that their composition played a central role in determining the constituents of planetary bodies. An important point on the composition of planetesimals is the initial rock/ice ratio because if enough volatiles were efficiently incorporated in the planetesimals, it implies a very different evolutionary track.

Once accretion was completed, it was thought that the radioactive components (mostly $^{26}$Al and $^{60}$Fe) produced enough internal heating to melt ices. The aqueous fluids generated interacted with primary phases to form the secondary minerals found in chondrites. Where insufficient radioactive heat was available, as found by Kunihiro et al. (2004), other processes such as shock compaction probably released water locally and over relatively short timescales (Trigo-Rodríguez et al., 2006; Trigo-Rodríguez and Blum, 2010; Rubin, 2012; Lindgren et al., 2015). The interaction of water with minerals typically involves elemental mobilization through dissolution, and precipitation of reaction products. This process is of paramount importance in some undifferentiated meteorites such as carbonaceous and ordinary chondrites (Jewitt et al., 2007).

The name chondrite originates from the rounded silicate spherules called chondrules, which together with CAIs are the main constituents of most chondrite classes apart for the most highly aqueously altered ones (i.e., some CM and all CI chondrites). Chondrules consist of refractory silicate minerals and opaque phases (e.g., metal and sulphide), and are dominant components of these meteorites that come from parent bodies that experienced hydrothermal activity (Dyl et al., 2012), with the exception again of CI chondrites (Brearley and Jones, 1998). These macroscopic components are enclosed by a fine-grained matrix containing micrometric or even nanometric particles, some of which originated in other stars; i.e., the so-called presolar grains. The porous nature of such fine-grained aggregates makes them potential places to preserve and retain ices and hydrated phases that were present in the outer disk (Abreu and Brearley, 2010; Moyano-Cambero et al., 2016; Stroud et al., 2016; Singerling and Brearley, 2017). There is evidence in the literature of the action of water in some of these components of chondrites, an aspect on which this review is focused.

Consequently, the chondrites are conglomerates of fine dust (a mixture of silicates, oxides, metal, sulphides and organic constituents), chondrules and CAIs of different sizes and proportions (Brearley & Jones, 1998). It was proposed that the chondrites came from undifferentiated bodies because the materials from which they formed are heterogeneous in chemical and isotopic composition. However, new paleomagnetic results are challenging this probably simplistic scenario with presumably larger parent bodies (Carporzen et al., 2011; Shah et al., 2017). In fact, many CI and CM chondrites have being metamorphosed to some extent (Rubin et al., 2007) and they could have experienced thermal annealing linked to post-impact shock (Rubin, 2004).

The chondrites comprise carbonaceous, ordinary, enstatite, Rumuruti (R) and anomalous (ungrouped) classes, which in turn are subdivided into 15 groups (Weisberg et al., 2006). Each group displays different degrees of thermal or aqueous alteration (see the reviews by Brearley & Jones, 1998; Dobrică and Brearley, 2014), with some of the thermal processing potentially taking place by solar heating at short perihelion distances (Marchi et al., 2009; Trigo-Rodríguez et al., 2009). Remarkably, all chondrites are described as chemically primitive because the ratios of their major, non-volatile elements (Fe, Si, Mg, Al, Ca, etc…) are close to those observed in the Sun (Anders and Grevese, 1989; Lodders, 2003). In particular, CI, CM, and

CR groups are those that experienced a higher degree of parent body aqueous alteration (Rubin et al., 2007; Abreu and Brearley, 2010; Trigo-Rodríguez et al., 2015). The CR group is recognized as being especially pristine and can record early solar system processes in between the fine-grained textures of their matrices (Abreu and Brearley, 2010; Moyano-Cambero et al., 2016). Many chondrite groups contain secondary minerals that indicate they experienced thermal metamorphism and aqueous processing (Table 1).

This review aims to provide additional insights into the presence of accretionary ice and hydrated phases in carbonaceous asteroids by describing and interpreting the evidence available in CM and CR carbonaceous chondrites. Our interpretations are compared with recent models and results in the scientific literature.

**2. Instrumental procedures.**

We have analyzed several carbonaceous chondrites with clear evidence for the action of water (Table 1). Our technical procedure is conventional with 20 μm thin sections of each meteorite studied microscopically using the following techniques. Mineral compositions were determined when further understanding of the effects of aqueous alteration in each meteorite was required.

Table 1 (see Page 27 for Tables)

2.1. Petrographic Microscope

At the clean meteorite laboratory of the Institute of Space Sciences we use a Zeiss Scope, with magnifications up to 500X, and with a Motic BA310Pol Binocular microscope. This technique was used to study properties such as undulatory extinction, mosaicism, pleocroism, interference colours (with crossed nicols), mineral microstructures, etc. It is an interesting way for starting to recognize principal features and minerals and enables a great deal of information if used accurately. With this instrument we can distinguish the action of water in some minerals as shown by the presence of aqueously formed minerals, for example phyllosilicates. Crystalline silicate phases become diffuse and with distinguishable alteration colors when transformed into serpentine or other phyllosilicates. In addition, we also sought evidence for lineations and fractures as they denote the existence of preferential shock compaction.

2.2. Scanning Electron Microscopy (SEM)

Samples were studied at the *Institut Català de Nanociència i Nanotecnologia* (ICN2) using a FEI Quanta 650 FEG with a Back Scattered Electron Detector (BSED). The Quanta 650 was operated in low-vacuum mode with each thin section uncoated, and EDS analyses were used to obtain X-Ray spectra and semi-quantitative composition from specific points. Secondary Electron (SE) and Back-Scattered Electron (BSE) imaging was used to study in detail sites of interest in the sample, with magnifications up to 30,000 X. The SEM was also used to select regions of interest (ROIs) for subsequent chemical analysis by electron microprobe and EDS.

2.3. Energy-Dispersive X-ray Spectroscopy (EDS)

We characterized the mineralogy of the thin sections by performing non-destructive EDS analysis by SEM. EDS was used to create elemental maps of the ROIs (Fig. 3). The EDS detector was an Inca 250 SSD XMax20 with Peltier cooling and with a detector area of 20 mm$^2$, attached to the SEM at the Institut Català de Nanociència i Nanotecnologia (ICN2).

2.4. Electron microprobe

BSE imaging and electron microprobe analyses were performed using a JEOL JXA-8900 electron microprobe equipped with five wavelength-dispersive spectrometers at the Universitat de Barcelona. The electron microprobe fires an electron beam at the sample, and detects the characteristic X-rays emitted by each element, when electrons from higher energy shells fill the vacancies left by the electrons scattered by the beam. The energy of the X-rays is characteristic of the difference in energy between two electron shells, and of the atomic structure of the element, which enables quantitative elemental analysis of the specimen.

2.5. Ultra High Resolution Transmission Electron Microscopy (UHRTEM).

In order to study the mineral structure of CM chondrites at the nanoscale, chips of two meteorites (Murchison and Cold Bokkeveld) that were ion thinned to electron transparency using a Fischione 1050 model Ar ion mill at CIC (Granada University). The samples were then studied using a FEI Titan G2 60-300 microscope at CIC with a high brightness electron gun (X-FEG) operated at 300 kV and equipped with a Cs image corrector (CEOS). Analytical electron microscopy (AEM) used a SUPER-X silicon-drift windowless EDX detector. The X-ray spectra were collected in STEM (Scanning Transmission Electron Microscopy) mode with imaging using a HAADF (High Angle Annular Dark Field) detector. Digital X-Ray maps were also collected on selected areas of the samples. For quantitative analyses the EDX data were corrected using the thin-film method (Cliff and Lorimer, 1975; Lorimer and Cliff, 1976) with K-factors determined using mineral standards: albite (Na, Al), anorthite (Al, Ca), anorthoclase (Na, Al), Augite (Mg, Al, Ca & Fe), biotite (Mg, Al, K, Fe), scapolite (Na, Al, Ca), spesartine (Al, Mn), hemimorphite (Zn), microcline (Al, K), muscovite (Al, K), olivine (Mg, Fe), rhodonite (Mn, Fe), titanite (Ca, Ti), and osumilite (Mg, Al, K, Fe). The resulting KAB factors were: Na 1.18 (0.03), Mg 1.10 (0.03), Al 1.00 (0.02), Si 1.00, K 1.12 (0.04), Ca 1.03 (0.03), Ti 1.28 (0.04), Mn 1.33 (0.01), Fe 1.37 (0.03) and Zn 1.53 (0.01). Atomic concentration ratios were converted into formulae according to stoichiometry (number of O atoms in theoretical formulae).

# 3. Results: Evidence of accretionary phases in CM and CR chondrites

This section focuses on describing evidence in CCs for accretion of hydrated phases. The best evidence is found in some CM and CR chondrites for which water availability was limited. Once the hydrated minerals had been incorporated into these meteorites, the water could have been released by moderate heating from three main sources: self-gravity, collisional compaction, and radiogenic decay. It is worth mentioning that self-gravity is an unlikely heat source because the CM chondrites probably had a parent body of few tens of km of diameter, but this process cannot be discounted completely.

As a consequence of aqueous alteration, minor phases grew in the pores of these chondrites. The alteration products in Murchison and other CM chondrites have been widely described in previous papers (e.g., Fuchs et al, 1973; Trigo-Rodríguez et al., 2006; Rubin et al., 2007). Here we focus on two CMs, namely Murchison and MET 01070, and two CR chondrites, namely Renazzo, and GRA 95229.

## 3.1. CM CHONDRITE GROUP

CM chondrites are breccias and aqueously altered rocks that contain up to ~9 wt. % indigenous water, mostly bound in phyllosilicates (Bischoff et al., 2006; Rubin et al., 2007). All CM chondrites have undergone extensive parent body aqueous alteration, and the products of these reactions obscure evidence for primordial accretionary processes. The porous matrices of most CM chondrites contain clumps of minerals formed by aqueous alteration, consisting mainly of phyllosilicate and tochilinite intergrown with pentlandite and Ni-bearing pyrrhotite. A full description of these phases is in Rubin et al. (2007).

Parent body heating was originally proposed to have been mild in the CM and CR groups, as the maximum inferred temperatures in both CC groups are 50 ºC and <150 ºC, respectively (Zolensky et al., 1993). Now we know that some of the CCs were thermally metamorphosed at temperatures of up to 800 ºC and above after aqueous alteration (Garenne et al., 2014; Doyle et al., 2015; King et al., 2017). Aqueous alteration produces changes in the silicates that are transformed into phyllosilicates (Velbel and Palmer, 2011, Velbel et al., 2012). This suggests that collisions also affected the bulk chemistry of their parent bodies, and probably participating to some extent in aqueous alteration and compaction of the different groups (Zolensky and McSween, 1988; Zolensky et al., 2008). An extended chronology for the setting of aqueous alteration has been given by Lee et al. (2012).

A new aqueous alteration sequence for CM chondrites has been proposed that provides evidence of progressive aqueous alteration, ranging from type 3.0 (pristine, unaltered materials) to type 2.0 (highly altered rocks, formerly classified CM1) (Rubin et al., 2007). Some alteration features in CMs result from processes that went to completion at the beginning of the alteration sequence, such as alteration of primary igneous glass in chondrules, production of aqueous altered minerals, and the formation of secondary sulphide grains. Other petrologic properties reflect processes active throughout the alteration sequence: formation of phyllosilicates, oxidation of metallic Fe-Ni, destruction of isolated matrix silicate grains, alteration of chondrule phenocrysts, changes in the composition of secondary minerals, and the precipitation of carbonates (Trigo-Rodríguez et al., 2006). Most CM chondrites have experienced different degrees of aqueous alteration, but some chemically linked ungrouped chondrite like e.g. Acfer 094 seems to have largely escaped it and have remained pristine (Rubin et al., 2007).

### 3.1.1. MURCHISON CHONDRITE

Murchison is one of the least aqueously altered CM chondrites, in common with Murray, but the extent of hydration of both meteorites is heterogeneous (Trigo-Rodríguez et al., 2006; Rubin et al., 2007; Trigo-Rodríguez et al., 2015). There is no doubt that the collisions had significant effects in the evolution of the parent body of CM chondrites (Trigo-Rodríguez et al., 2006; Rubin, 2012). The petrographic texture is complex, and Murchison is a breccia (Rubin et al., 2007). Murchison has type-I chondrules with tochilinite inclusions, metal grains and it has many spherical troilite and pyrrhotite grains in its matrix. Tochilinite appears as a single phase but it exhibits variable composition, and is correlated with P-bearing sulfides, which are associated with pyroxene and forsterite (Palmer and Lauretta, 2011). Even though tochilinite and P-bearing sulfides are the common phases, tochilinite also contains chromium-rich phases

Metal grains in Murchison often have aureoles that are composed mainly of oxides and formed by corrosion (Hanowski and Brearley, 2000). Aqueous alteration minerals also occur in the fine-grained matrix and include sulphides and carbonates that precipitated from water (Trigo-Rodríguez et al., 2006; Rubin et al., 2007; Lee et al., 2014). It seems likely that most chondrules accreted into the parent body with a fine-grained porous mantle that probably hosted hydrated minerals or even dirty ices (Figure 1). The secondary minerals preferentially grew where the pores were available within chondrule mantles and the surrounding matrix, and the variable sizes of crystals and textures probably reflects the presence of volatile phases that disappeared under the action of moderate heat and aqueous processing, probably associated with shock compaction (Trigo-Rodríguez and Blum, 2010). Radiogenic heating and collisional compaction are processes that probably produced moderate heat that helped evaporation of the volatiles, but leaving aqueous alteration minerals when the components in solution precipitated into the pores of the meteorite. The way in which the aqueous alteration proceeded from highly porous progenitors retaining volatiles in their interior has been described in detail (Trigo-Rodríguez et al., 2006; Rubin et al., 2007).

Figure 1 (see page 28 for Figures)

Aqueous alteration of Murchison chondrules is not pervasive, with some areas being more highly altered than others (Lee and Lindgren, 2016). This is nicely shown in Figures 2 and 3. Figure 2 is a backscattered electron (BSE) image of a chondrule whose rim is diffuse, apart for on the left hand side (square inset region, Fig. 2a,c). Fig. 2c,d illustrates a corroded metal grain contained in the chondrule. It is also possible to follow a line of Fe-oxides that penetrates into the chondrules, which can be seen particularly well in Fig. 3a. Some S and Ca are also associated with such aqueous alteration minerals produced in the interior of the chondrules (sulphides and carbonates are common as well).

Figures 2, and 3

Our previous nanoscale study of Murchison identified a chemically unequilibrated matrix that is highly heterogeneous under HRTEM examination (Trigo-Rodríguez et al., 2017). Micrometer-sized windows of Murchison matrix display characteristic layering of phyllosilicates associated with highly reactive metal grains and sulphides. These minerals exemplify an extraordinary accretionary diversity in this meteorite and highlight the need to study other highly unequilibrated chondrites to confirm the existence of hydrous mineral phases in the protoplanetary disk (see e.g. Fegley and Prinn, 1989; Lodders and Fegley, 2011). We think that this nanoscale complexity may be indicative of the formation conditions of this meteorite, and the minimal thermal processing that occurred in its parent asteroid.
We have additionally investigated CM2 chondrites to seek evidence for the incorporation of hydrated phases and ices into the parent body, which may be best identified using nanoscale observations b UHRTEM (Trigo-Rodríguez et al., 2017). To highlight the level of aqueous alteration of the matrix we include here an image of Murchison matrix obtained using the technique. The defined areas in Figure 4 correspond to: 1) Troilite (FeS), 2) Talc, 3-8) Mg-rich serpentine, 9) calcic amphibole, and 10) calcic pyroxene. The bright spots in the Fe+S images are clearly troilite that are adjacent to phyllosilicates that have suffered significant parent body alteration. In fact, Brearley (1997) described that clinoenstatite have been altered to amphibole and talc (two of the identified minerals in our ROIs #2 and #9) nearby contraction cracks in the case of CV3 carbonaceous chondrite Allende. It is well known that parent body aqueous alteration usually leads to a progressive loss of Mg-rich anhydrous silicates because the Mg goes into the fluid and precipitates to form Mg-rich phyllosilicates (Trigo-Rodríguez et al., 2006; Rubin et al., 2007).

Probably due to extensive parent body aqueous alteration no clear evidence for pre-accretionary hydrated phases in carbonaceous chondrites has been published. Our working hypothesis is that extensive aqueous alteration often predates the oldest evidence of accretionary hydrated minerals. The CM2 chondrite Murchison is particularly interesting because it is a breccia that contains regions with different degrees of aqueous alteration, probably due to limited water availability. In this sense, we found that some ROIs of Murchison have escaped extensive interaction with water (Rubin et al., 2007). Less altered regions contain highly reactive phases along with phyllosilicates, sulphides and oxides (see Fig. 4, and Trigo-Rodríguez et al., 2017). It is difficult to understand the co-existence of reactive minerals (like e.g. FeS) with hydrated phases at this fine scale. As a consequence, we envision that these minerals were probably stacked together in the aggregate structure, and scarcely affected by extensive parent body aqueous alteration. If we are correct, the relative absence of evidence for accreted hydrated minerals could be direct consequence of the bias imposed when we study CCs at microscale. We should undertake more studies of pristine meteorites at the nanoscale given that theoretical predictions and observations of presolar grains and IDPs indicate that the minerals available in the protoplanetary disk were nanometric (Zinner, 2003; Lodders and Amari, 2005; Lodders and Fegley, 2011). We have noticed that the matrix of Cold Bokkeveld, the other CM2 studied using UHRTEM, is much more highly altered (Trigo-Rodríguez et al., 2017). In consequence, given the different degree of aqueous alteration found in CM chondrites (Rubin et al., 2007), only these experiencing moderate parent body alteration (like e.g. Murchison) could be of use in the search of pre-accretionary hydrated minerals.

In view of current evidence it is difficult to make any definitive statements about wet accretion, but we think this should not be ruled out without additional studies of the carbonaceous chondrites at nanometre scale. Probably using existing techniques to study these rocks at the nanoscale evidence of pre-accretionary hydrated minerals could be obtained. We hope that our interpretation of the matrix mineralogy of Murchison can promote further research into the possibility that carbonaceous chondrites contain nanoscale evidence of wet accretion.

Figure 4.

### 3.1.2. MET 01070 CM2 CHONDRITE

We have chosen the CM2.0 MET 01070 because it has suffered pervasive aqueous alteration, and exhibits long lenses probably caused by precipitation from an aqueous fluid (see Fig. 5 and Rubin et al., 2007). The lenses are formed of aqueous alteration products, range appreciably in thickness, and surround large chondrule pseudomorphs. We conclude that these lenses were produced when a water-rich front with various elements in solution precipitated phases containing Ca, P, Ni, and S. As consequence of this process, Ca-carbonate (mainly $CaCO_3$) and Ca-phosphate grains were produced within the lens. In order to form this precipitation front, a substantial volume of water was needed within the parent body perhaps due to a thermal gradient or to compaction. Other phases within the lens are Ni-bearing sulfide (e.g. pentlandite), which are present in the meteorite matrix (Table 2).

Figure 5.

These kind of extended lenses are not usual in CMs, and are clearly a consequence of extensive aqueous alteration (Rubin et al., 2007). The significant aqueous alteration differences observed in specimens of the most pristine CM and CRs suggest that water availability was probably limited for most of them (see Trigo-Rodríguez, 2015 and references therein). These observations associated with pristine chondrites are of key relevance in the search for evidence of hydrated minerals or ices accreted from the protoplanetary disk. We predict that the accretion of such water-rich phases should produce distinctive aqueous alteration typically being local and nearly static in nature.

## 3.2. CR CHONDRITE GROUP

This group of chondrites is of key interest with regards to the evolution of volatile-rich species in the protoplanetary disk. The CR group has large chondrules, ranging from 0.8 to 4.4 mm in diameter, with an abundance of 40 to 60 vol % (Hutchison, 2004). These chondrules are rich in forsterite ($Mg_2SiO_4$), enstatite ($MgSiO_3$) and Fe,Ni-metal rich inclusions. They have conspicuous rims of metals and silicates, which could have been formed by processes taking place either within the protoplanetary disk or during accretion. The fine-grained matrix is dark and optically opaque and comprises up to 30-51 vol % of the meteorites. It is mainly composed of olivine, but also contains carbonates, pyrrhotite ($Fe_{1-x}S$; $x = 0$ to $0.2$), which is a variant of troilite (FeS), pentlandite ($(Ni,Fe)_9S_8$), magnetite ($Fe_3O_4$), serpentine ($(Mg,Fe)_3Si_2O_5(OH)_4$) and chlorides (Brearley and Jones, 1998; Abreu and Brearley, 2010). CAIs are rare (<3 vol %) but where present are rich in melilite ($(Ca,Na)_2(Al,Mg,Fe^{2+})$) and spinel ($MgAl_2O_4$) and also contain silicates and other oxides. CR chondrites also contain dark phyllosilicate rich inclusions (<8 vol %), Fe-Ni metal inclusions (5-8 vol %) and sulfides (1-4 vol %) (Trigo-Rodríguez, 2012).

The CRs contain minerals that have experienced much less parent body aqueous alteration than the CMs so that pre-accretionary processes **could be** more apparent. For this reason we have selected this chondrite group for our discussion related to hydration processes in the protoplanetary disk. In general, the CR chondrites are composed of chondrules, fine-grained matrix, and usually tiny CAIs and ameboid-olivine aggregates (AOAs). Type-I forsteritic chondrules rich in Fe-Ni metal grains compise about ~96% of the chondrule population, type-II chondrules constitute ~3 %, while other types are minor (Schrader et al., 2013). The matrices of CR chondrites consist of abundant submicron Fe-rich hydrated amorphous silicate grains, mixed with nanometer-sized phyllosilicates and organics (Le Guillou and Brearley, 2014; Le Guillou et al., 2015; Alexander et al., 2017). These widespread oxidized and hydrated amorphous silicates provide clues of the alteration conditions and $H_2$ degassing of asteroids. In the following sub-sections some interesting CR specimens are described in the context of possible evidence of post-accretionary aqueous alteration.

### 3.2.1. RENAZZO

Renazzo is a pristine CR chondrite that has avoided extensive parent-body thermal metamorphism. As a consequence its metal grains have escaped recrystallization, and are not mixed as kamacite and taenite phases (Wasson and Rubin, 2010). It is a nice example of meteorite that has undergone minor aqueous alteration, clearly much less than has been experienced by most of the CM chondrites described in previous section. Moreover, this meteorite was recovered soon after its fall, so that it has not undergone terrestrial alteration. For these reasons Renazzo has clues to processes of post-accretional aqueous alteration that could be associated with direct modification of minerals by water derived from accreted hydrated phases or ices. To find definitive evidence in that regard we need to study the fine-grained matrix that probably retained water in these two forms, as documented for LAP 02342 CR chondrite (Moyano-Cambero et al., 2016; Stroud et al., 2016)

Petrographic evidence in support of the aforementioned scenario is that Renazzo exhibits features characteristics of static aqueous alteration (see e.g. Trigo-Rodríguez et al., 2006). This type of alteration occurs when water is scarce, and is highlighted by Fig. 6. Chondrule metal grains in the left hand side of the image have clear evidence of partial corrosion, as well as some minerals present near the chondrule's contact with the matrix. Mobilization of Fe, S and Ni has formed some tiny precipitates in the surrounding matrix, near the chondrule itself.

Figure 6.

Such features have been described previously in other CR3 chondrites (Abreu and Brearley, 2010), and these authors emphasized the importance of this chondrite group for understanding the first phases of planetesimal accretion. The matrices of some CR chondrites often have significant FeO-enrichments in comparison to the less altered ones, again indicating that the alteration could have operated at local scale and for a limited time. These observations are consistent with static aqueous alteration in the parent asteroid, so it probably requires the incorporation of water ice in the matrix of CCs. Then, water availability assured its incorporation to the structure of the minerals through chemical bonds, preserving it of escaping.

### 3.2.2. GRAVES NUNATAKS 95229

This selected meteorite is an altered CR2, and quite remarkable for its extraordinary abundance of amino acids. Most CR chondrites are breccias*, and all are petrological type 2. Chondrules in GRA 95229 are large (<2 mm), with an abundance of 50 to 60 vol %. They are mainly composed of Fe-poor olivine and pyroxene. Chondrule rims are rare, poorly defined and often discontinuous (Abreu and Brearley, 2008). They are rich in Fe sulfides and magnetite (Brearley and Jones, 1998). Surrounding matrix regions are rich in silicates, with magnetite, rare Fe-Ni sulfides and chondrule fragments. GRA 95229 contains a large proportion of organic materials, e.g. amino acids like glycine, alanine and isovaline, in comparison with other CR chondrites. They are also more abundant than in almost all other primitive chondrites, such as CM2s Murchison and Murray, which have one of the larger organic matter proportions of all chondrites (Martins et. al., 2007).

Extensive aqueous alteration is evident in other CR chondrites (Trigo-Rodríguez, 2015). Water played a major role in mobilizing specific elements and altered several mineral phases to produce carbonates, oxides and sulphides. In some cases the action of water was pervasive, particularly with regards to metal and troilite grains, and participated in the complete replacement of mineral grains located in the matrix. Pyrrhotite was replaced during aqueous alteration of the CR2 chondrites Graves Nunataks (GRA) 95229 and Elephant Morraine (EET) 92159. After such replacement the voids were filled by submicron-sized magnetite grains. In some cases the replacement is incomplete and some pyrrhotite remains. Interestingly, such growth of magnetite is expected to occur from the precipitation of Fe from a low-temperature aqueous solutions (see e.g. Fig. 7). A similar action of water on pyrrhotite grains has been observed in the Kaidun meteorite (Zolensky & Ivanov, 2003; Trigo-Rodríguez et al., 2013).

Fig. 7

3.2.3. La Paz 02342

The LAP 02342 CR chondrite highlights how the Antarctic collection continued to provide new specimens of extraordinary cosmochemical interest. LAP 02342 has probably experienced a lower degree of aqueous alteration than other CR chondrites, and we have identified some clear signatures of accretion of ices in different components (Moyano-Cambero et al., 2016; Stroud et al., 2016; Nittler et al., 2019). This meteorite contains an unusual ~100 µm diameter highly C-rich clast that is composed of a fine-grained mixture of isotopically anomalous organic matter, $^{16}$O-poor Na-sulfates, crystalline and amorphous silicate grains (including GEMS - Glass with Embedded Metal and Sulfide), sulfides, and abundant presolar grains. The Na-sulfates surrounding the micro-xenolith are isotopically similar to the rare cosmic symplectite, likely reflecting the isotopic composition of a primordial water reservoir that accreted as ice within the clast (Nittler et al., 2019). This ultra-carbonaceous clast may represent a cometary building block that accreted into the matrix of the CR chondrite parent asteroid. These observations are also consistent with hydrogen and nitrogen isotopic anomalies in organic matter retained in the fine-grained matrices of CCs that were probably inherited from other unsampled bodies of cometary nature present in the protoplanetary disk at the time of formation of carbonaceous chondrites (Busemann et al., 2006)

These pristine CR chondrites can also provide new clues about protoplanetary disk chemistry. LAP 02342 hosts a chondrule with a primordial S-rich mantle (Figure 8) altering the surrounding matrix (see e.g. Moyano-Cambero et al., 2016). It has been recently found that S chemistry was significant in protoplanetary disks (Fuente et al., 2016). It is important to note that if such chemistry existed, primordial ices could be enriched in this element because S-bearing molecules can be slightly polar: $S_2H$, $H_2S$, $H_2S_2$, SH, etc. so it is likely that they could easily attach to ices, and be accreted into ice-rich chondrule rims. According to these observations, and other isotopic observations of the rim described in Nittler et al. (2019), we found that this S-rich mantle is evidence for ice accretion in the chondrule rim. We interpreted the accretional signatures because, once on the CR host matrix, the sublimation of the ice promoted the formation of the observed $^{16}$O-poor Na-sulfates (Nittler et al., 2019).

**Figure 8**

## 4. Discussion

### 4.1 Using the extent of aqueous alteration to understand pre-accretionary hydration

Parent bodies of the samples described here are carbon-rich asteroids that are highly unequilibrated mixtures of different materials, some crystalline and created close to the Sun, others formed at greater heliocentric distances. Most of these bodies accreted in the outer protoplanetary disk in the presence of ices, organics and hydrated mixtures. The most primitive materials coming from these bodies are very different to terrestrial rocks because they are chemically highly unequilibrated (Brearley & Jones, 1998). The rock-forming materials building the so-called transitional asteroids are probably closer to the materials forming comets, thus invoking a continuum between asteroid and cometary materials (Briani et al., 2011; Trigo-Rodríguez, 2015). The NASA Stardust mission confirmed that comet 81P/Wild 2 is also composed of micrometre-size refractory mineral grains blown by the stellar wind to their formation regions (Brownlee et al., 2006). Laboratory experiments (Blum et al., 2006) suggest that the comets were highly porous, so that they probably retained volatiles and hydrated phases during the early stages of Solar System formation. The relative absence of hydrous phases in grains collected during the Stardust mission was probably a consequence of heating during the capture process, which led to vaporization of the most volatile phases (Trigo-Rodríguez et al., 2008).

Thermochemical modeling suggests that some minerals formed in the protoplanetary disk were hydrated, and probably water was incorporated as ice- and carbon-rich aggregates (Lodders, 2003; Ebel, 2006). Despite expectations, evidence for the accretion of hydrated phases and ices in the CM and CR chondrites has remained elusive owing to overprinting by thermal metamorphism and aqueous alteration (Bischoff, 1998). Once the parent bodies were formed as aggregates (Blum et al., 2006), these volatile phases were compacted during collisions, and sublimed or liquefied to saturate the rock. For that reason it is very important to identify pristine CCs, and recent studies on CR chondrites are promising (Abreu and Brearley, 2010; Trigo-Rodríguez et al., 2015; Moyano-Cambero et al., 2016).

To gain insight on the amount of water incorporated into the parent bodies of carbonaceous chondrites, Howard et al. (2015) quantified the modal abundances of major mineral phases (with abundances >1 wt.%) using Position Sensitive Detector X-rayDiffraction (PSD-XRD). They concluded that the variability between CCs in their measured hydration indicates that either accretion of ices was heterogeneous, or fluid was widely mobilized. Obviously, most meteorite samples are biased samples because they come from bodies that experienced collisions sufficiently energetic to launch the ejecta into heliocentric orbits (Beitz et al., 2016). Despite this current bias, which should be mitigated by future sample-return missions, Howard et al. (2015) found using PSD-XRD that the initial mass fraction of $H_2O$ inside of their parent asteroids was <20 wt.%. This suggests a relatively small fraction of hydrated phases and/or ices were incorporated, probably as minor constituents of meteorite matrices (Trigo-Rodríguez, 2015). Recent H-isotope investigations evidence also suggest two different sources of accreted water in the CM chondrites (Piani et al., 2018). Most secondary minerals in CCs show O-isotopic signatures that are consistent with accretion of a source of water in the inner Solar System that became $^{16}$O-rich, while water related with outer minor bodies has the $^{16}$O-poor signature (Marrocchi et al., 2018; Nittler et al., 2019). This fact has been explained because Jupiter acted as a barrier precluding significant inward transport of outer solar system materials, with the probable exception of cometary fragments produced by tidal disruption during close approaches to the giant planets. An example for such contribution could the $^{16}$O-poor Na-rich sulfates in the border of the C-rich micro-xenolith discovered by Nittler et al. (2018).

The study of primordial fayalite grains can be very relevant to understand the complex accretion scenario as was exemplified in Marrocchi et al. (2018) where the origin of water in CCs was investigated in CV chondrites using O isotopes hosted in several minerals. They conclude that the process controlling the O-isotopic composition of CV chondrites was related to the isotopic equilibrium between $^{16}$O-rich anhydrous silicates and a $^{17}$O and $^{18}$O-rich fluid.

Other aqueous alteration minerals including calcite are common in the different chondrite groups (Table 1), but the action of water produced distinctive mineralogy, which depends on each chondrite group, temperature and differences in the availability of liquid water. Our results clearly show that these often tiny mineral grains preferentially grew in the matrix pores (Trigo-Rodríguez et al., 2006; Lee et al., 2012, 2014). Most of the aqueous alteration seems to be static, short-time events that produced distinctive features like the aureoles in Murchison that formed around native metal grains by oxidation (Hanowski and Brearley, 2000). Calcite and other complex carbonates are also useful in dating hydration processes as their $^{53}$Mn-$^{53}$Cr ages precise formation ages for minerals in the CI and CM chondrites (for an update see e.g. Fujiya et al., 2013). Results indicate that hydration of the CI parent body occurred between 3 and 9 Ma after formation of the CAIs, themselves dated to 4.567 Ga (Amelin et al., 2002). Lee et al. (2012) found that some CM chondrites, for example Queen Elizabeth Range (QUE) 93005, have shorter alteration periods starting about 4 Ma after the formation of the oldest solar system solids. A recent study of CM2 carbonaceous chondrite Lonewolf Nunataks (LON) 94101 reveals that it is possible to decipher the existence of multiple carbonate generations from intrinsic oxygen isotope differences (Lee et al., 2012; 2013).

A formation scenario for the Murchison CM2 chondrite involves a transitional asteroid that probably accreted both anhydrous and hydrous materials. Radiogenic heating contributed at an early stage to releasing part of the water that altered some materials, but only locally because the water was inhomogeneously distributed. Later on, because of successive impacts of the parent asteroid with other bodies a brecciated structure was created, the porosity of the matrix decreased and all the materials were significantly compacted (Trigo-Rodríguez et al., 2006). Chondrules and inclusions were tough objects that by their nature resisted compaction. For this reason it was in matrix material where the collisions had their greatest effects. Over time Murchison's materials started to fragment and comminute. Some primordial mantles were detached and some aqueous alteration could act locally as a consequence of water release if direct evaporation did not occur. Due to all this collisional gardening most of the known CM chondrites are *breccias* of materials with diverse alteration or metamorphic histories, but finally built by impact processes (Hanna et al., 2015). As a consequence of multiple impacts experienced by chondritic asteroids since their formation (see e.g. Beitz et al., 2016), pore spaces collapsed and the materials were sheared, heated and shattered. Evidence for this scenario is the discovery of free mantles in the matrices of CM chondrites (Trigo-Rodríguez et al. 2006) that, even when probably they were accretionary in origin, also experienced compaction during shock events (Bland et al., 2014). These impacts could have promoted the release of liquid water, initiating aqueous alteration of some minerals and a mobilization of some elements that gave room for growth of secondary (or altered) minerals like these fine-grained aqueous alteration phases. To describe and date these processes is extremely important in an astrobiological perspective because aqueous alteration processes could have directly contributed to catalyse the extremely diverse organic species found in carbonaceous chondrites (Rotelli et al., 2016)

On the other hand, radiogenic heating of chondritic parent bodies was size-dependent and constrained during the first 10 Ma after their formation. We know this from the carbonates, whose mean size varies with the extent of aqueous alteration. The highly altered Alais and Tonsk CI meteorites have large carbonate grains (Endreβ & Bischoff, 1996). This process is probably associated with the precipitation of minerals including carbonates, sulphides and phosphides in the matrices of these meteorites. These bodies accreted live $^{53}$Mn that was mobilized by water and later precipitated into carbonates, preferentially in the void spaces of the highly porous parent bodies (Trigo-Rodríguez *et al.*, 2006; Rubin *et al.*, 2007). We found that dolomite and complex carbonates are developed at the expense of Ca-carbonate in the highly altered CM chondrites (Rubin *et al.*, 2007; Lee et al., 2014). Extensive aqueous alteration could also have promoted aqueous flow in the CM parent body producing the lenses identified in MET 01070 (Trigo-Rodríguez & Rubin, 2006; Rubin *et al.*, 2007, see Fig. 5). It has been also proposed that the collapse of matrix pore spaces occurred due to shock propagation associated with impact compaction of the CM and CV parent bodies (Trigo-Rodríguez et al., 2006; Rubin, 2012). In such circumstances water could have being vaporized quite quickly leaving the aqueous alteration fronts observed in MET 01070.

4.2 Atomistic pictures of material/water interfaces relevant for the CCs

The present work aims to provide proof for the presence of hydrous phases and their action during the accretion of CCs in the outer regions of the protoplanetary disk. In this section we synthesise computational chemistry results with quantum mechanical simulations that can provide atomic-scale information on the interaction of water with materials present in CCs. Results can in turn can shed some light to the initial steps of water alteration in some minerals present in CCs.

Some of the clearest evidence for aqueous alteration is the presence/formation of phyllosilcates in CCs. Phyllosilicates are also referred to as hydrous silicates because they incorporate water or hydroxyls. Our recent work on water incorporation has used quantum chemical calculations (Rimola & Trigo-Rodríguez, 2017), with forsterite ($Mg_2SiO_4$) being used as a silicate test case. In that work, we studied the interaction of 12 water molecules with the (010), (001) and (110) surfaces of $Mg_2SiO_4$ crystals, which are extended planes in the $Mg_2SiO_4$ crystal structure (as well as with an amorphous surface). It is worth mentioning that these different $Mg_2SiO_4$ surfaces have contrasting stabilities. The stability of a surface is correlated with its reactive behavior, and specifically the less stable a surface the greater its reactivity. The different $Mg_2SiO_4/(H_2O)_{12}$ systems were geometrically optimized to reach a stable structure of minimum energy. Figure **9** shows the results obtained. Irrespective of the surface, water molecules interact with the outermost surface $Mg^{2+}$ cations through their O atoms and also via hydrogen-bond interactions between the O surface atoms of forsterite and the H protons of water. Moreover, these interactions only occur in the first water layer, whereas the second water layer was found to be engaged by a H-bond network. However, it was found that the resulting $Mg^{2+}/H_2O$ interaction depends on the stability of the silicate surface. For the most stable surfaces, (010) and the (001), water molecules interacting directly with $Mg^{2+}$ are molecularly adsorbed on the cation; that is, they do not split into $OH^+$ H. By contrast, for the less stable the (110) surfaces and the amorphous surfaces those water molecules interacting with the $Mg^{2+}$ surface cation dissociate whereby one H proton of water is transferred to a nearby O surface atom, thereby resulting in the formation of Mg-OH and Si-OH surface groups. Similar results (namely, water dissociation upon $Mg_2SiO_4$ adsorption) were also found by Priggiobe et al. (2013), in which the $Mg_2SiO/H_2O$ interface was studied by increasing progressively the number of water molecules to adsorb. This water dissociation on the most reactive $Mg_2SiO_4$ surfaces can be interpreted as the first step for the transformation of silicates into phyllosilicates. In our particular case, formation of chrysotile can take place by hydration of forsterite; i.e., $2Mg_2SiO_4 + 3H_2O \rightarrow Mg_3Si_2O_5(OH)_4 + Mg(OH)_2$.

Continuing with the same idea that phyllosilicates are clear evidence for aqueous alteration during the accretion of CCs, the literature contains a set of computational chemistry studies based on ab initio molecular dynamic simulations essentially focused on the dehydroxylation of two phyllosilicates: pyrophillite (Molina-Montes et al., 2008a; Molina-Montes et al., 2008b; Molina-Montes et al., 2010) and smectite (Muñoz-Santiburcio et al., 2012; Muñoz-Santiburcio et al., 2016). These authors identified different competitive paths for the dehydroxilation of these hydrous minerals, but the major conclusion is that in order to carry out the reaction high temperatures ($\approx$1500 K) are required because of the high activation energies (about 60 kcal mol$^{-1}$ for pyrophillite and 50 kcal mol$^{-1}$ for smectite). This result indicates that once it has taken place, phyllosilicate dehydroxylation (namely, reverting the reaction) requires extreme conditions. If this is not the case, then the phyllosilicate remains as a stable material.

The results related to the dissociation of water upon surface interaction also suggest that in some CCs the presence of iron oxides (and particularly magnetite, $Fe_3O_4$) is evidence for alteration by water. We stated above (Section 3.2.2) that iron sulphides, and in particular pyrrhotite (a nonstoichiometric form of FeS), can be replaced by magnetite during aqueous alteration. Several theoretical works address the interaction of water with iron sulphide surfaces. Stirling et al. (2003) studied the interaction of water with the (100) surface of pyrite ($FeS_2$). Results of this work indicated that the most stable $FeS_2/H_2O$ structure is that in which all the water molecules are molecularly adsorbed on the outermost Fe cations. However, the authors also identified a metastable structure in which some water molecules are in their dissociated state as a consequence of a proton transfer from $H_2O$ interacting with Fe to a S atom, thus resulting in the formation of Fe-OH and S-H surface groups (see Figure 10a). Remarkably, dissociated water is stabilized by the H-bond interactions established with the rest of water molecules present in the interface. In absence of water in excess, the dissociative state is unstable and evolves towards molecular water. Another interesting work is that of Dzade et al. (2016), where the adsorption of water on a clean and oxygen-covered (011) surface of mackinawite (a FeS-type mineral) was studied. Results showed that, on the clean surfaces (namely, in absence of other co-adsorbents), water dissociation is endothermic from its molecularly adsorbed state (see Figure 10b). However, in the presence of a preadsorbed oxygen atom on a Fe site (namely, oxygen-covered surface), water dissociation is feasible since a spontaneous proton transfer reaction from water to the preadsorbed O atom was observed, thus forming two $Fe^{3+}$-$OH^-$ ferric hydroxide groups (see Figure 10c). The fact that $H_2O$ can be dissociated upon Fe interaction opens up a route towards the formation of iron oxides in the presence of water. Indeed, in common with silicates, water dissociation on iron sulphides is a way to incorporate O into the mineral structure, leading to its oxidation. Obviously, taking into account only the water dissociation process we cannot explain the full iron sulphide → iron oxide conversion process. Other intriguing aspects remain to be answered such as removal of the surface H atoms (originating from water), the growth of the iron oxide, and the need or otherwise of co-adsorbents activating the process. This water dissociation phenomenon, however, allows us to determine the initial steps towards the formation of iron oxides from iron sulphides.

Another interesting point is the mobilization of certain elements due to the action of water. Water can play a major role in the diffusion and migration of elements (particularly in the form of cations such as $Na^+$), thus altering the composition of the CC materials. A nice theoretical work demonstrating the capability of water in mobilizing metal cations is that of Mignon et al. (2010). These authors used ab initio molecular dynamics simulations to explore the fate of $Li^+$, $Na^+$ and $K^+$ bound on the surfaces of the interlayer regions on montmorillonite clays under wet conditions. Results indicated that $K^+$ cations remain attached to the interlayer surfaces. In contrast, $Li^+$ cations are hydrated by four water molecules, thus trapping them and removing them from the interlayer surfaces. $Na^+$ cations show an intermediate behavior: whether or not they become hydrated depends on the position they occupy in the interlayer region. (Mignon et al., 2010) found a correlation of their results with the water-cation and surface-cation affinities. That is, the binding affinity of the alkali cations with water followed $Li^+ > Na^+ > K^+$, while their interaction with the interlayer surfaces follow the sequence of (from less to more favorable) $Li^+ < Na^+ < K^+$. These results provide atomic-scale evidence for the potential role of water in trapping and sequestering certain metal cations placed on the outermost positions of the materials, thus favoring their mobility and migration to other positions as clear effect of water alteration.

## 5. Conclusions

In view of current evidence, the CCs contain highly reactive minerals that have being continuously reaching the terrestrial surface, although the flux has significantly changed over time (Trigo-Rodríguez et al., 2017). The study of the highly unequilibrated materials forming these meteorites reveals that they formed part of undifferentiated parent bodies accreted in the outer protoplanetary disk, and their composition reinforce the idea of a continuous between asteroids and comets. These bodies formed from primordial protoplanetary disk materials, and at the very beginning were subjected to planetary perturbations, collisions with other bodies and fragmentations during close approaches to planets so probably they were easily disrupted (Trigo-Rodríguez and Blum, 2009; Trigo-Rodríguez, 2015). The delivery of these materials to terrestrial planets increased during the Late Heavy Bombardment (Gomes et al., 2005). Consequently, it seems plausible that at early times and subsequent ulterior periods of time the Earth was subjected to a meteoritic flux at least 5-6 orders of magnitude greater than the current one (Trigo-Rodríguez et al., 2004) that has been estimated to be ~40,000 Tm/year (Brownlee, 2001). Consequently, we predict that a large amount of chondritic materials reached the early Earth's surface at an annual rate of thousands of billions of metric tons. Hence the amount of volatiles and catalytic minerals delivered under such meteoroid high-flux circumstances was also very significant, probably playing a key role in fertilizing the Earth's surface (Rotelli et al., 2016). As consequence of our present study we have reached the following conclusions:

- We have presented evidence for aqueous alteration in carbonaceous chondrites that can be explained by the incorporation of hydrated materials into the fine-grained matrix. As their parent bodies were compacted by collisional processing and self-gravity, only those that escaped significant metamorphism or aqueous alteration can provide direct clues needed to understand the delivery of water to terrestrial planets.

- Many carbonaceous chondrites contain valuable chemical, isotopic and mineralogical information on the nebular environment, but exhibit densities and porosities different to their precursor materials. The effects of aqueous alteration and impact metamorphism make finding undifferentiated bodies preserving primordial physical properties a difficult task, but meteorites such as ALHA77307 and Acfer 094 shows that they do exist (Rubin *et al.,* 2007; Trigo-Rodríguez & Blum, 2009a). Sample-return missions from the future exploration of ice-rich bodies stored among the Centaurs or Kuiper Belt populations will provide new clues and answers.

- Carbonaceous chondrites are the legacy of the first accretionary stages of our protoplanetary disk. Accretion of abundant organic- and ice-rich phases occurred behind the so-called snow line, so in the outer disk distinctive materials were available to form transitional bodies. The recognized progenitors of these primitive meteorites are volatile-rich asteroids, and comets. To explain the current O-isotopic evidence, our current formation scenario indicates that Jupiter precluded significant inward transport of outer disk comets, being probably the cause for the two distinguishable sources of water identified so far. The outer border of the inner disk where the CCs mostly incorpored $^{16}$O-rich water, while the $^{16}$O-poor signature seems to be characteristic of outer disk water ice hosted by cometary materials.

- *Transitional* bodies probably accreted not only anhydrous minerals, but also significant amounts of ice, hydrated minerals, and organic materials that made them highly porous in the very beginning. These materials accreted as very small grains, or attached to other phases forming meteorite matrices. Ices and organics could have helped to bind the primordial dust aggregates without the need for significant compaction.

- Carbonaceous chondrites formed from outer disk aggregates containing variable ice/rock ratios. Radiogenic elements (e.g. $^{26}$Al, $^{60}$Fe) were incorporated from the stellar environments from which our planetary system formed. Then, once consolidated, radiogenic decay probably provided the heat required to release enough water for aqueous alteration during the first 10 Ma after accretion. Mineralogical evidence, mostly formation and dating of carbonates, indicates that aqueous alteration and outgassing transformed these bodies at that early stage, thus being consistent with radiogenic decay.

- The brecciated nature, deformation and lineation of some components contained in CCs suggest that shock produced by stochastic collisions was an additional source of heat that, due to the relatively limited abundance of water, mostly promoted static aqueous alteration.

- Atomistic simulations based on computational chemistry methods are potential powerful tools for providing atomic-scale insights into water-mediated alteration of minerals present in CCs. The simulations provide a comprehensive understanding of the first steps for the conversion of anhydrous silicates into phyllosilicates, the oxidation (namely, incorporation of O) of iron sulfides, and the mobility of metal cations present at the mineral surfaces

Concerning future work we consider that it is urgent to perform radiogenic dating of some more of the aqueous alteration products in order to reconstruct the sequence of events involved in the alteration of the parent bodies of CM and CR chondrites. Some meteorites have being studied, but still we lack measurements dating these pre-accretionary minerals described in this work. These studies are of key importance to establish the pathway of delivery of volatiles to the terrestrial planets.

**Acknowledgements**

We thank two anonymous reviewers that improved significantly this manuscript. Spanish Ministry of Science and Innovation under research projects AYA2015-67175-P and CTQ2017-89132-P are acknowledged, and we also thank the UK Science and Technology Facilities Council for funding through project ST/N000846/1. Mike Zolensky is acknowledged for kindly providing the Murchison and Renazzo pristine sections studied in this work. AR is indebted to "Ramón y Cajal" program. ST made this study in the frame of a PhD. on Physics at the Autonomous University of Barcelona (UAB). M. del Mar Abad is acknowledged by her interpretation of the data obtained of Murchison CM2 using the HR-TEM image (Fig. 4) obtained by JMTR at Centro de Instrumentación Científica (CIC), Universidad de Granada.

# REFERENCES

Abreu N.M. and Brearley, A.J. Geoch. Cosm. Acta 74, 1146 (2010)

Alexander, C.M.O'D., McKeegan, K.D. and Altwegg, K. Space Sci. Rev. 214: 36 (2018)

Alexander, C.M.O'D., G.D. Cody , B.T. De Gregorio , L.R. Nittler , and R.M. Stroud. Chemie der Erde 77, 227 (2017)

Amelin, Y., Krot, A.N., Hutcheon, I.D., & Ulyanov, A.A. Science, 297, 1678 (2002)

Anders, E. and Grevese, N. Geoch. Cosmoch. Acta, 53, 197 (1989)

Armitage P.J. Annu. Rev. Astron. Astrophys., 49, 195 (2011)

Beitz, E., Güttler, C., Nakamura, A. M., Tsuchiyama, A. & Blum, J. Icarus, 225, 558 (2013)

Beitz, E., Blum, J., Parisi, M. G., and Trigo-Rodriguez, J. Ap.J. 824:1, article id. 12 (2016)

Bischoff, A. Meteorit. Planet. Sci. 33, 1113 (1998)

Bischoff, A., Scott, E.R.D., Metzler, K. & Goodrich, C.A. In *Meteorites and the Early Solar System II*, D.S. Lauretta & H.Y. McSween, eds. (The University of Arizona Press, Tucson, 2006), pp.679-712

Bland, P. A., Collins, G. S., Davison, T. M., Abreu, N. M., Ciesla, F. J., Muxworthy, A. R., and Moore, J. Nature Communications 5, id. 5451 (2014)

Blum, J., Schräpler, R., Davidson, B.J.R. & Trigo-Rodríguez, J.M. Ap J. 652, 1768 (2006).
Boss A.P. Ap.J. 764, 194 (2013)

Brearley, A.J. Science 276, 1103-1105 (1997)

Brearley, A. J. and Jones, R.H. In *Planetary Materials*, J.J. Papike, ed., *Reviews in Mineralogy*, 36, Mineralogical Society of America, Washington, 1998), pp-1-398 (1998)

Brearley A. J. In Meteorites and the Early Solar System II, ed. D. S. Lauretta and H. Y. McSween, (Univ. Arizona Press, Tucson, 2006), pp. 587-624 (2006)

Briani, G., Morbidelli, A., Gounelle, M., and Nesvorný, D. Meteoritics & Planetary Science 46, 1863 (2011)

Browning L., McSween H. and Zolensky M. Geochim. Cosmochim. Acta 60, 2621 (1996)

Brownlee D.E. In *Accretion of Extraterrestrial Matter Throughout Earth's history*. B. Peucker-Ehrenbrink and B. Schmitz (eds.), Kluwer Academic/Plenum Publishers, New York, USA, pp. 1-12 (2001)

Brownlee D., et al. Science 314, 1711 (2006)

Busemann, H. Young, A.F., Alexander, C.O'D., Hoppe, P., Mukhopadhyay, S., Nittler, L.R.. Science 312, 727 (2006).

Carporzen L., Weiss, B.P., Elkins-Tanton, L.T., Shuster, D.L., Ebel, D. and Gattacceca, J. Proc. National Academy Sciences 108, 6386 (2011)

Cliff G. and Lorimer G.W. J. Microscopy 103, 203 (1975)

Dobrică, E., Brearley, A. J. Meteoritics & Planetary Science, 49, 1323 (2014)

Doyle, P.M., Jogo, K., Nagashima, K., Krot, A.N., Wakita, S., Ciesl, F.J., and Hutcheon, I.D. Nature Communications, doi: 10.1038/ncomms8444 (2015)

Dzade, N.Y., Roldan, A.. de Leeuw, N.H. (2016) J. Phys. Chem. C, 120, 21441−21450.

Dyl, K.A., Bischoff, A., Ziegler, K., Young, E.D., Wimmer, K. and Bland, P.A. Proceedings of the National Academy of Sciences, 109, 18306 (2012)

Endreβ, M. & Bischoff, A. Geochimica & Cosmochimica Acta, 60, 489 (1996)

Fegley, B., Jr. and Prinn, R.G. In *The Formation and Evolution of Planetary Systems*, Space Telescope Science Institute Symposium Series, H.A. Weaver and L. Danly (eds.) (1989)

Fuchs L. H., Olsen E., and Jensen K. J. Smithsonian Contributions to Earth Science 10:39 (1973)

Fuente, A., Cernicharo, J., Roueff, E., Gerin, M., Pety, J., Marcelino, N., Bachiller, R., Lefloch, B., Roncero, O., and Aguado, A. Astronomy and Astrophysics 593, A94 (2016)

Garenne, A., Beck, P., Montes-Hernandez, G., Chiriac, R., Toche, F., Quirico, E., Bonal, L., and Schmitt, B. Geochimica et Cosmochimica Acta 137, 93 (2014)

Gomes R., Levison H. F., Tsiganis K., Morbidelli A. Origin of the cataclysmic Late Heavy Bombardment period of the terrestrial planets. Nature 435, 466 (2005)

Hanna, R.D., Ketcham, R.A., Zolensky, M., Behr, W.M. Geochimica et Cosmochimica Acta 171, 256 (2015)

Hanowski N. P. and Brearley A. J. Meteorit. Planet. Sci. 35, 1291 (2000)

Howard, K.T., Alexander, C.M.O'D, Schrader, D.L., and Dyl, K.A. Geochim. Cosmochim. Acta 149, 206 (2015)

Hutchison R. Meteorites, Cambridge Univ. Press, 506 pp. (2004)

Jewitt, D., Chizmadia, L., Grimm, R. and Prialnik, D. In Protostars and Planets V, B. Reipurth, D. Jewitt and K. Keil (eds.), Univ. Arizona Press, pp. 863-867 (2007)

King, A.J., Schofield, P.F., and Russell, S.S. Meteoritics and Planetary Science 52, 1197 (2017)

Le Guillou, C., and Brearley, A.J. Geochimica et Cosmochimica Acta 131, 344 (2014)

Le Guillou, C., Changela, H.G., Brearley, A.J. (2015) Earth and Planetary Science Letters 420, 162 (2015)

Kunihiro T., Rubin, A.E., McKeegan, K.D. and Wasson, J.T. Geochimica Cosmochimica Acta 68, 3599 (2004)

Lee, M.R., Lindgren, P., Sofe, M.R., O'D Alexander, C.M. & Wang, J. Geochimica Cosmochimica Acta, 92, 148 (2012).

Lee, M.R. and Lindgren, P. Meteoritics and Planetary Science 51, 1003 (2016)

Lee, M.R., Lindgren, P. and Sofe, M.R. Geochimica et Cosmochimica Acta 144, 126 (2014)

Lindgren, P., Hanna, R.D., Dobson, K.J., Tomkinson, T. and Lee, M.R. Geochimica et Cosmochimica Acta 148, 159-178. (2015)

Lodders, K. The Astrophysical Journal 591, 1220 (2003)

Lodders, K. and Fegley, B. In *Chemistry of the Solar System*, RSC Publishers, ISBN: 978-0-85404-128-2, 496 pp. (2011)

Lodders, K. and Amari S. Chemie der Erde 65, 93–166 (2005)

Lorimer G.W. and Cliff G. (1976) In Electron Microscopy in Mineralogy, Ed. by H. R. Wenk, Berlin Springer-Verlag, Berlin, 506 (1976)

Marchi, S., Delbó, M., Morbidelli, A., Paolicchi, P., and Lazzarin, M. Monthly Notices of the Royal Astronomical Society 400, 147 (2009)

Marrocchi, Y., Bekaert, D.V. and Piani, L. Earth Planet. Sci. Lett. 482, 23 (2018)

Martins, Z., Alexander, C. M. O'D., Orzechowska, G. E., Fogel, M. L., and Ehrenfreund, P. Meteoritics & Planetary Science 42, 2125 (2007)

Mignon, P., Ugliengo, P., Sodupe, M., Hernandez, E.R. (2010): Ab initio molecular dynamics study of the hydration of $Li^+$, $Na^+$ and $K^+$ in a montmorillonite model. Influence of isomorphic substitution. *Phys. Chem. Chem. Phys.*, 12, 688–697.

Molina-Montes, E., Donadio, D., Hernández-Laguna, A., Sainz-Díaz, C.I., Parrinello, M. (2008a): DFT Research on the Dehydroxylation Reaction of Pyrophyllite 1. First-Principle Molecular Dynamics Simulations. *J. Phys. Chem. B*, 112, 7051–7060.

Molina-Montes, E., Donadio, D., Hernández-Laguna, A., Sainz-Díaz, C.I., (2008b): DFT Research on the Dehydroxylation Reaction of Pyrophyllite 2. Characterization of Reactants, Intermediates, And Transition States along the Reaction Path. *J. Phys. Chem. A*, 112, 6373–6383.

Molina-Montes, E., Donadio, D., Hernández-Laguna, A., Sainz-Díaz, C.I., (2010): Exploring the Rehydroxylation Reaction of Pyrophyllite by Ab Initio Molecular Dynamics. *J. Phys. Chem. B*, 114, 7593–7601.

Moyano-Cambero, C. E., Nittler, L. R., Trigo-Rodríguez, J. M., Alexander, C. M. O'D., Davidson, J., and Stroud, R. M. 47th Lunar and Planetary Science Conference, LPI Contribution No. 1903, p.2537 (2016)

Muñoz-Santiburcio, D., Kosa, M., Hernández-Laguna, A., Sainz-Díaz, C.I., Parrinello, M. (2012): Ab Initio Molecular Dynamics Study of the Dehydroxylation Reaction in a Smectite Model. *J. Phys. Chem. C*, 116, 12203−12211.

Muñoz-Santiburcio, D., Hernández-Laguna, A., Sainz-Díaz, C.I. (2016): Simulating the Dehydroxylation Reaction in Smectite Models by Car−Parrinello-like-Born-Oppenheimer Molecular Dynamics and Metadynamics. *J. Phys. Chem. C*, 120, 28186−28192.

Nittler L.R., Trigo-Rodríguez, J.M. Stroud, R.M., De Gregorio, B.T., Alexander, C.M.O'D., Davidson, J., Moyano-Cambero, C.E., and Tanbakouei, S. Nature Communications, submitted, (2019)

Piani, L., Yurimoto, H. and Remusat, L. (2018) Nature Astronomy 2, 317 (2018)

Prigiobbe, V., Suarez Negreira, A., Wilcox, J. (2013): Interaction between Olivine and Water Based on Density Functional Theory Calculations. *J. Phys. Chem. C*, 117, 21203–21216.

Rimola, A., Trigo-Rodríguez, J.M. (2017): Atomistic Simulations of Aqueous Alteration Processes of Mafic Silicates in Carbonaceous Chondrites. Pp. 103-127 in: *Assessment and Mitigation of Asteroid Impact Hazards* (J.M. Trigo-Rodríguez, M. Gritsevich & H. Palme, editors). Astrophysics and Space Science Proceedings 46, Springer, Cham, Switzerland.

Rotelli L., Trigo-Rodríguez, J.M., Moyano-Cambero, C.E., Carota, E., Botta, L., Di Mauro, E., y Saladino R., The key role of meteorites in the formation of relevant prebiotic molecules in a formamide/water environment. Nature Sci. Rep. 6, Article number: 38888 (2016)

Rubin, A.E. Meteoritics and Planetary Science 32, 231 (1997)

Rubin, A.E. Geochimica Cosmochimica Acta, 68, 673 (2004)

Rubin, A.E. Geochimica Cosmochimica Acta, 90, 181 (2012)

Rubin, A., Trigo-Rodríguez, J.M., Huber, H. & Wasson, J.T. Geoch. Cosm. Acta, 71, 2361 (2007)

Schrader D.L., Connolly Jr. H.C., Lauretta D.S., Nagashima K., Huss G.S, Davidson J., Domanik K.J. Geoch. Cosmoch. Acta, 101, 302 (2013)

Schulz R., Hilchenbach, M., Langevin, Y., Kissel, J., Silen, J. et al. Nature 518, 216 (2015)

Shah, J., Bates, H.C., Muxworthy, A. R., Hezel D.C., Russell, S.S., and Genge M.J. Earth and Planetary Science Letters 475, 106 (2017)

Singerling, S.A. and Brearley, A.J. Meteoritics and Planetary Science, doi: 10.1111/maps.13108, 29 pp. (2018)

Stirling, A., Bernasconi, M., Parrinello, M. *J. Chem. Phys.*, 118, 8917 (2003)

Stroud, R. M., Nittler, L. R., Moyano-Cambero, C. E., Trigo-Rodriguez, J. M., Davidson, J., De Gregorio, B. T., and Alexander, C. M. O'D. 79th Annual Meeting of the Meteoritical Society, LPI Contribution No. 1921, id.6360 (2016)

Takir, D., Emery, J.P., McSween, H.Y., Hibbitts, C.A., Clark, R.N., Pearson, N., Wang, A. Meteoritics & Planetary Science 48, 1618 (2013)

Trigo-Rodríguez, J.M. In *Planetary Materials*, Edited by M.R. Lee and H. Leroux, 301 pp., ISBN:978-0903056-55-7, Published by the European Mineralogical Union and the Mineralogical Society of Great Britain and Ireland, p. 67 (2015)

Trigo-Rodríguez, J.M. and Blum, J. Planetary and Space Science 57, 243 (2009)

Trigo-Rodríguez, J.M., Rubin, A.E. and Wasson, J.T. Geoch. Cosmoch. Acta, 70, 1271 (2006)

Trigo-Rodríguez, J. M.; Llorca, J.; Oró, J. In *Life in the Universe: From the Miller Experiment to the Search for Life on Other Worlds*. Edited by J. Seckbach, J. Chela-Flores, T. Owen, and F. Raulin. 387 pp., ISBN 1-4020-2371-5 (HB), Published by Springer, Berlin, 2004, p.201 (2004)

Trigo-Rodriguez, J. M., Delbó, M., and Blum, J. European Planetary Science Congress 2009, held 14-18 September in Potsdam, Germany, p. 520 (2009)

Trigo-Rodríguez, J.M., Moyano-Cambero, C.E., Mestres, N., Fraxedas, J., Zolensky, M.E., Nakamura, T. & Martins, Z. In 44th Lunar and Planetary Sciences Conference, abstract #1929 (2013)

Trigo-Rodriguez, J. M.; Vila-Ruaix, A.; Alonso-Azcárate, J.; Abad, M. M. In Highlights on Spanish Astrophysics IX, Proceedings of the XII Scientific Meeting of the SEA, S. Arribas et al. (eds.), pp. 531-542 (2017)

Velbel M. A. and Palmer E. E. Clays and Clay Minerals 59, 416 (2011)

Velbel M. A., Tonui E. K., and Zolensky M. E. Geochimica et Cosmochimica Acta 87, 117 (2012)

Wasson, J.T., and Rubin A.E. Geoch. Cosmoch. Acta, 74, 2212 (2010)

Weisberg M.K., McCoy T.J. & Krot A.N. In *Meteorites and the Early Solar System II*, D.S. Lauretta & H. Y. McSween, eds.. (The University of Arizona Press, Tucson, 2006) pp. 19-52.

Zhang K, Blake G.A., and Bergin E.A. Astrophysical Journal Lett. 806, art. Id. L7, 6 pp. (2015)

Zinner E. In *Treatise on Geochemistry*, Volume 1. Editor: Andrew M. Davis. Executive Editors: Heinrich D. Holland and Karl K. Turekian. ISBN 0-08-043751-6. Elsevier, p.17-39 (2003)

Zolensky, M.E. & McSween H. Jr. In: Meteorites and the Early Solar System (J.F. Kerridge & M.S. Matthews, editors). pp. 114–143. The University of Arizona Press, Tucson, AZ, USA (1988).

Zolensky, M.E., Barret, T. & Browning, L. Geochimica Cosmochimica Acta, 57, 3123 (1993)

Zolensky, M. E. & Ivanov, A. Chemie der Erde, 63, 185 (2003)

Zolensky, M.E., Krot, A.N., Benedix, G. Record of low-temperature alteration in asteroids. Pp. 429–462 in: Oxygen in the Solar System (G.J. MacPherson, D.W. Mittlefehldt, J.H. Jones & S.B. Simon, editors). Reviews in Mineralogy & Geochemistry, 68. Mineralogical Society of America, Washington, D.C., (2008)

# TABLES

**Table 1.** Secondary minerals identified in the carbonaceous chondrites

| CI | CM | CR | CV/CK | Ungrouped | Mineral Group |
|---|---|---|---|---|---|
| Serpentine Saponite | Serpentine Chlorite Vermiculite | Serpentine Saponite | Serpentine Chlorite Micas | Serpentine Saponite | **Phyllosilicates** |
| Magnetite | Magnetite | Magnetite | Magnetite | Magnetite | **Oxides** |
| Calcite Dolomite Breunnerite Siderite | Calcite Dolomite Aragonite | Calcite | - | - | **Carbonates** |
| Apatite Merrilite | - | - | - | - | **Phosphates** |
| Pyrrhotite Pentlandite Cubanite Sulfur | Pyrrhotite Pentlandite Tochlinite Awaruite | Pyrrhotite Pentlandite | Pyrrhotite Pentlandite | - | **Sulphides** |
| - | Brucite Tochilinite | - | - | - | **Hydroxides** |
| - | Halite | - | - | - | **Halides** |

See also Rubin (1997), Trigo-Rodríguez *et al.* (2006) and Takir *et al.* (2013)

Table 2. Bulk composition of the aqueously formed lens in the MET 01070 CM2 chondrite compared with two Murchison aureoles (for more details see Rubin et al., 2007)

| Component | % in MET 01070 lens (n=23) | % in Murchison aureoles (n=2) |
|---|---|---|
| $SiO_2$ | 27.0 ± 4.6 | 5.61 |
| $TiO_2$ | 0.10 ± 0.06 | 0.16 |
| $Al_2O_3$ | 2.5 ±0.38 | 2.0 |
| FeO | 35.8 ±3.5 | 43.6 |
| $Cr_2O_3$ | 0.42 ±0.18 | 6.1 |
| MnO | 0.25 ±0.04 | 0.23 |
| MgO | 16.4 ±2.2 | 6.7 |
| CaO | 0.15 ±0.12 | 0.3 |
| $Na_2O$ | 0.20 ±0.03 | 0.23 |
| $K_2O$ | <0.04 ±0.01 | - |
| $P_2O_5$ | <0.04 ±0.04 | - |
| NiO | 0.72 ±0.44 | 7.6 |
| S | 1.5 ±0.97 | 8.3 |

# FIGURES

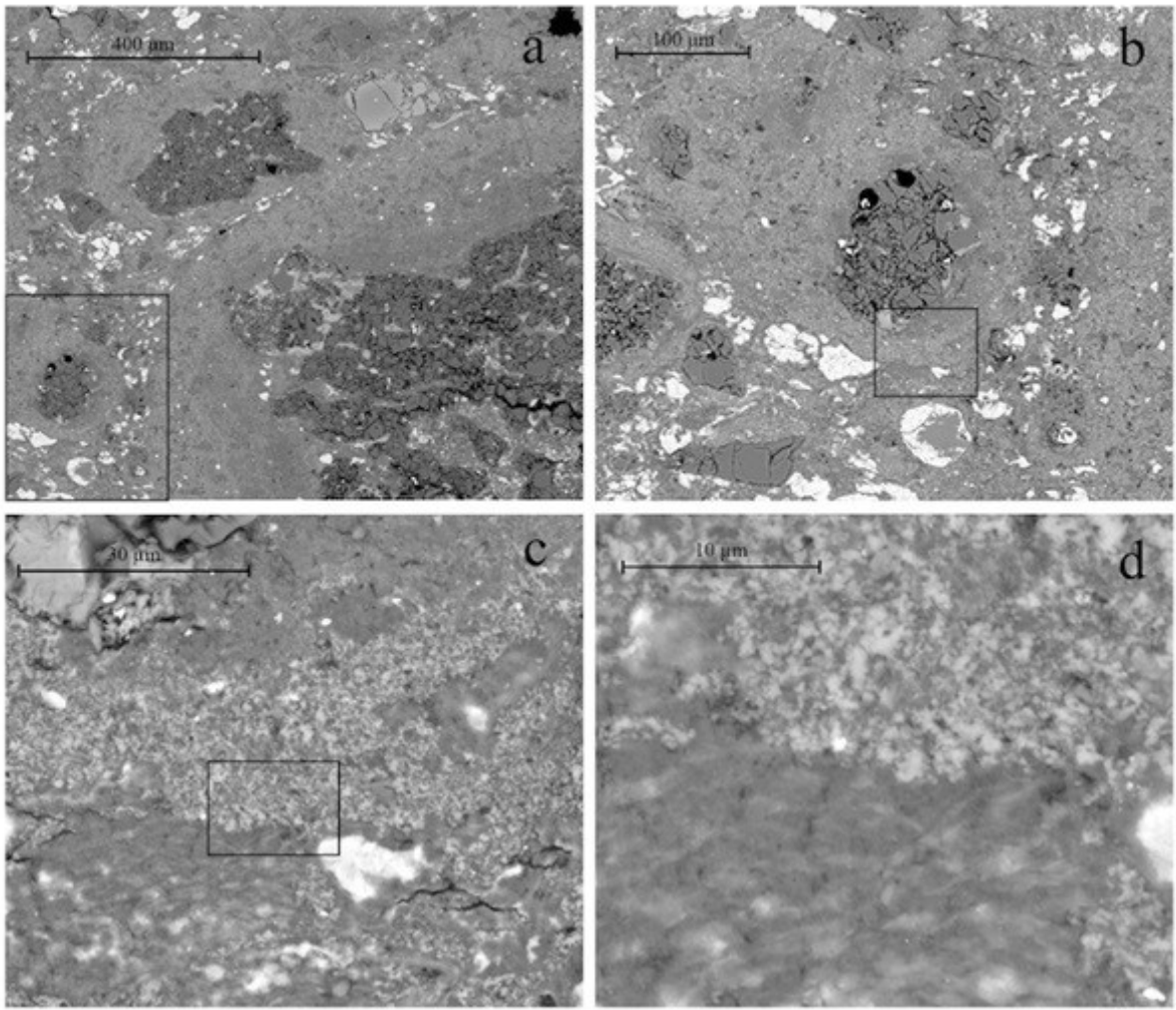

Figure 1. A zoom-in sequence of BSE images of a thin section of the Murchison CM2 chondrite that focuses on three chondrules and the texture of the mantles and surrounding matrix. a) Image showing the fine matrix texture, together with chondrules, and fragments. b) The lower left chondrule exhibits its own mantle and the different texture is highlighted at higher magnifications in c), and d).

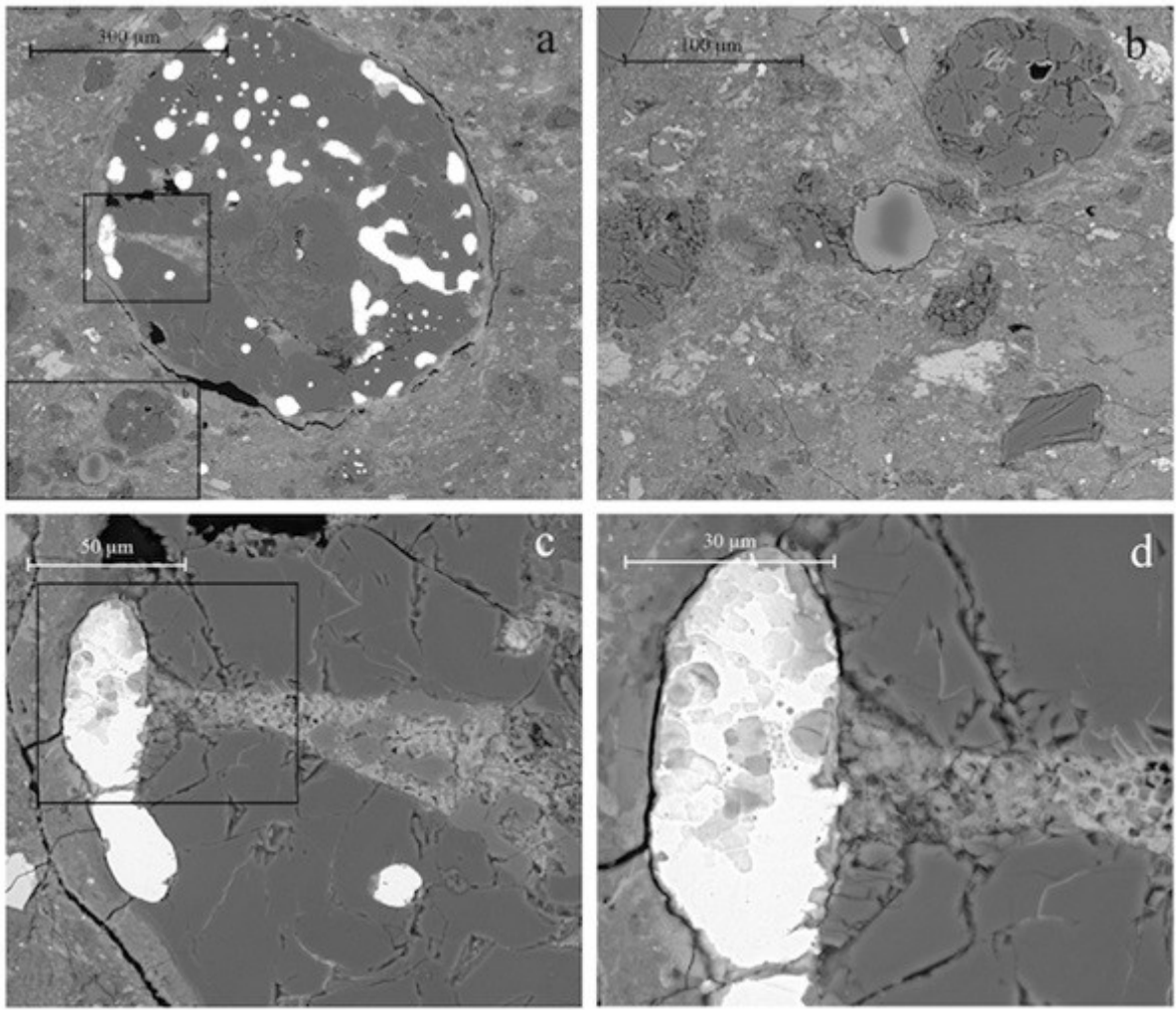

Figure 2. BSE images of a chondrule in the Murchison CM2 chondrite. (a) The metal-rich chondrule with two areas of interest framed. In (b) there is a relict grain while in (c) and (d) the metal grain is magnified in order to show how water corroded it and penetrates inside the chondrule.

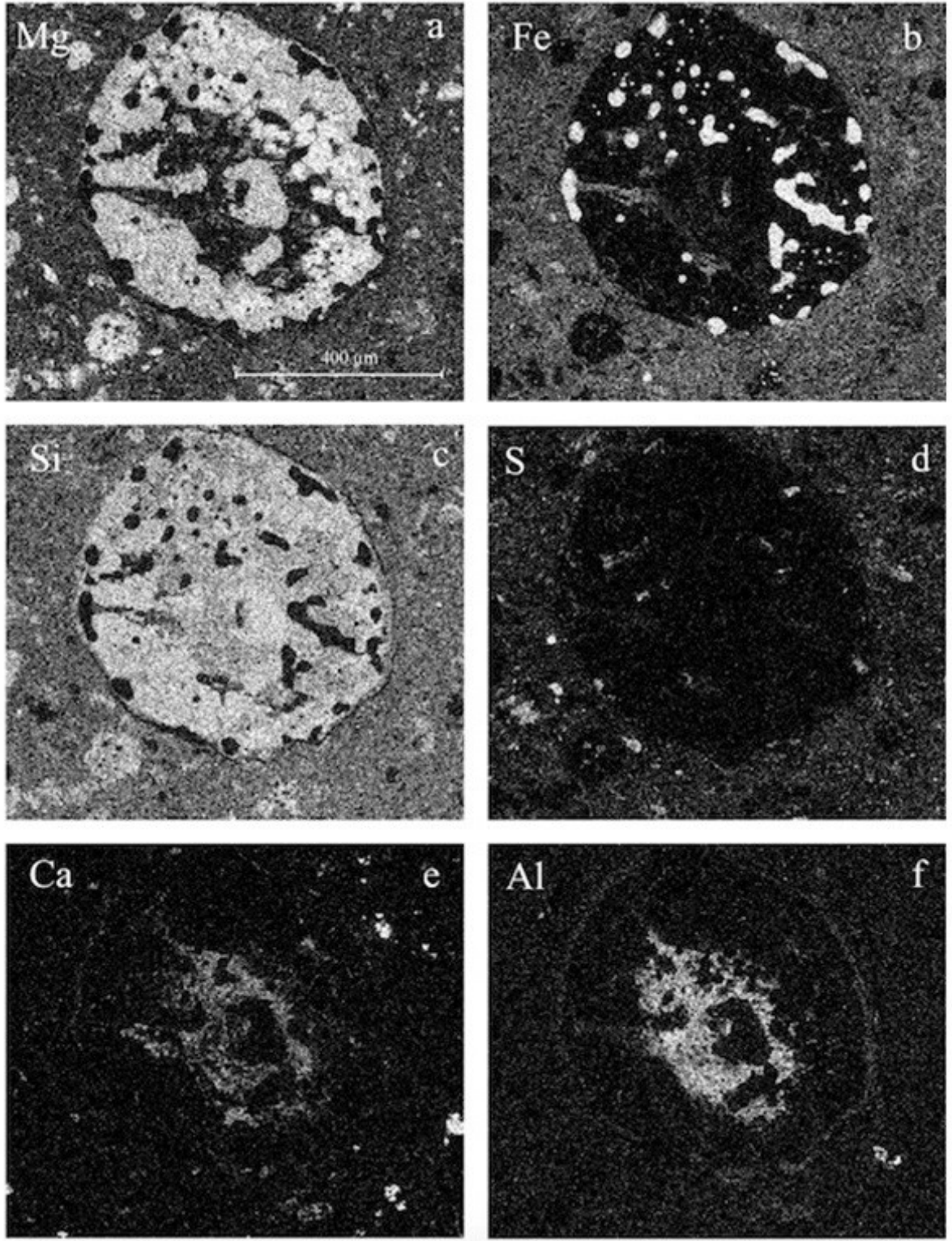

Figure 3. X-ray images of the chondrules described in Fig. 2 where the relative abundance of Mg, Fe, Si, S, Ca and Al in the rock-forming minerals is a function of their brightness.

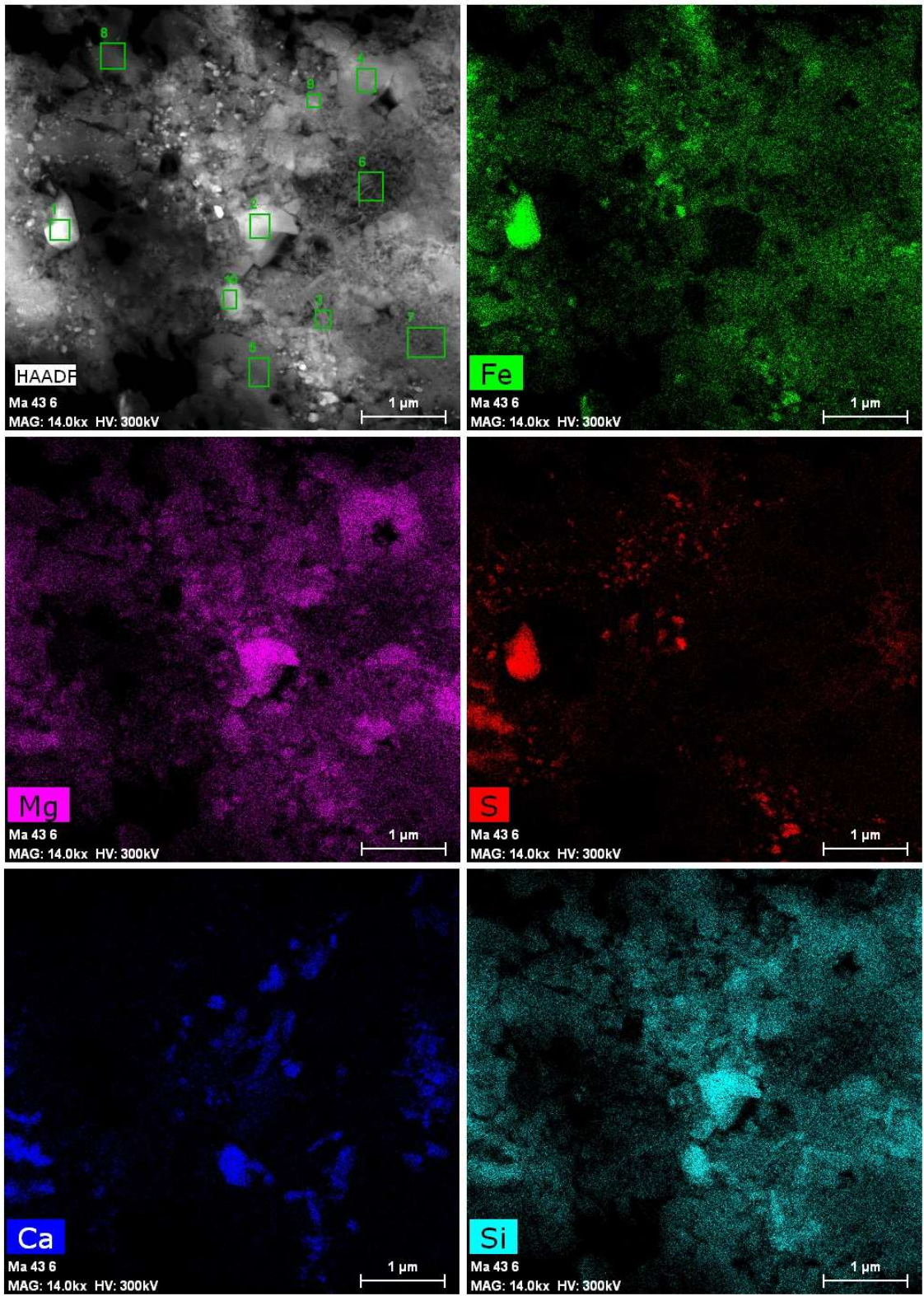

Figure 4. High-angle annular dark-field imaging (HAADF) and X-Ray elemental maps of Murchison matrix. For more details read the text.

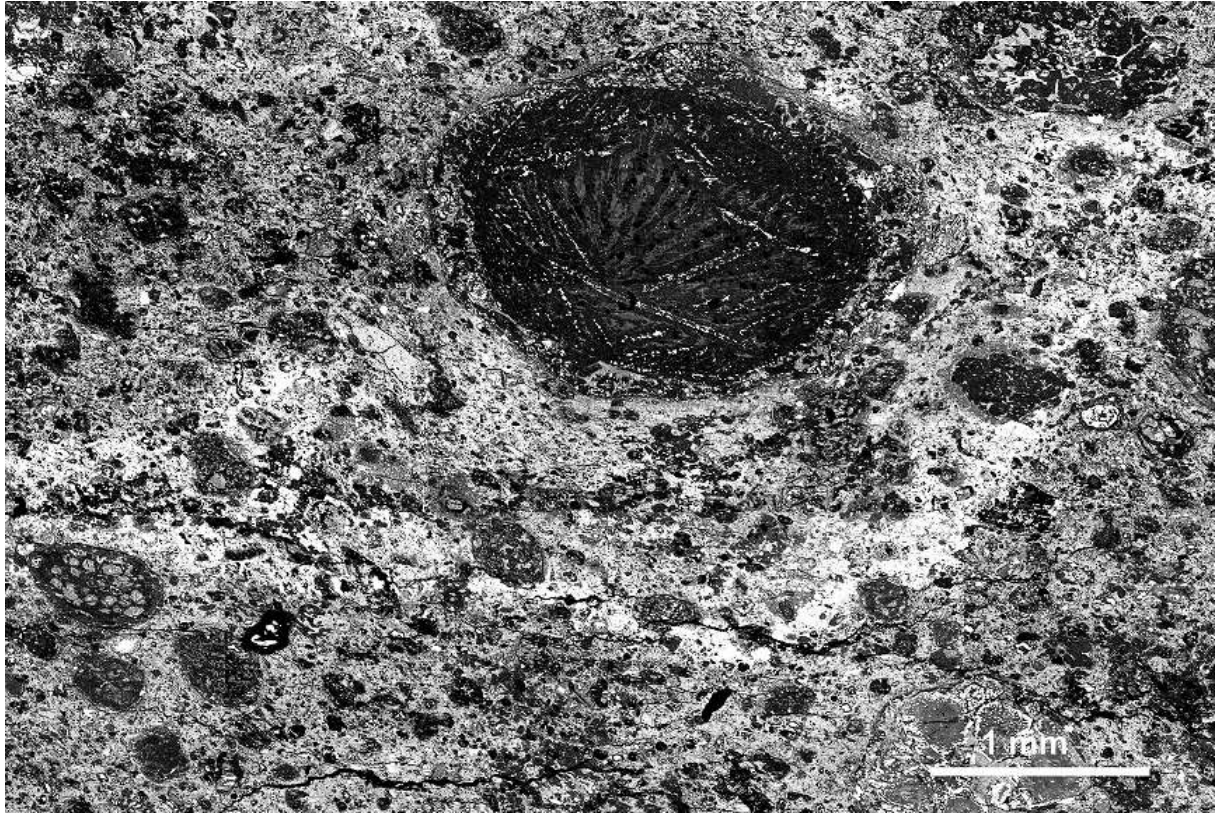

Figure 5. BSE image of a cm-sized lens found in the MET 01070 CM2.0 chondrite. The lens contains substantial amounts of fine-grained Ni-bearing sulphide and clumps of pentlandite grains (Trigo-Rodríguez & Rubin, 2006; Rubin et al., 2007).

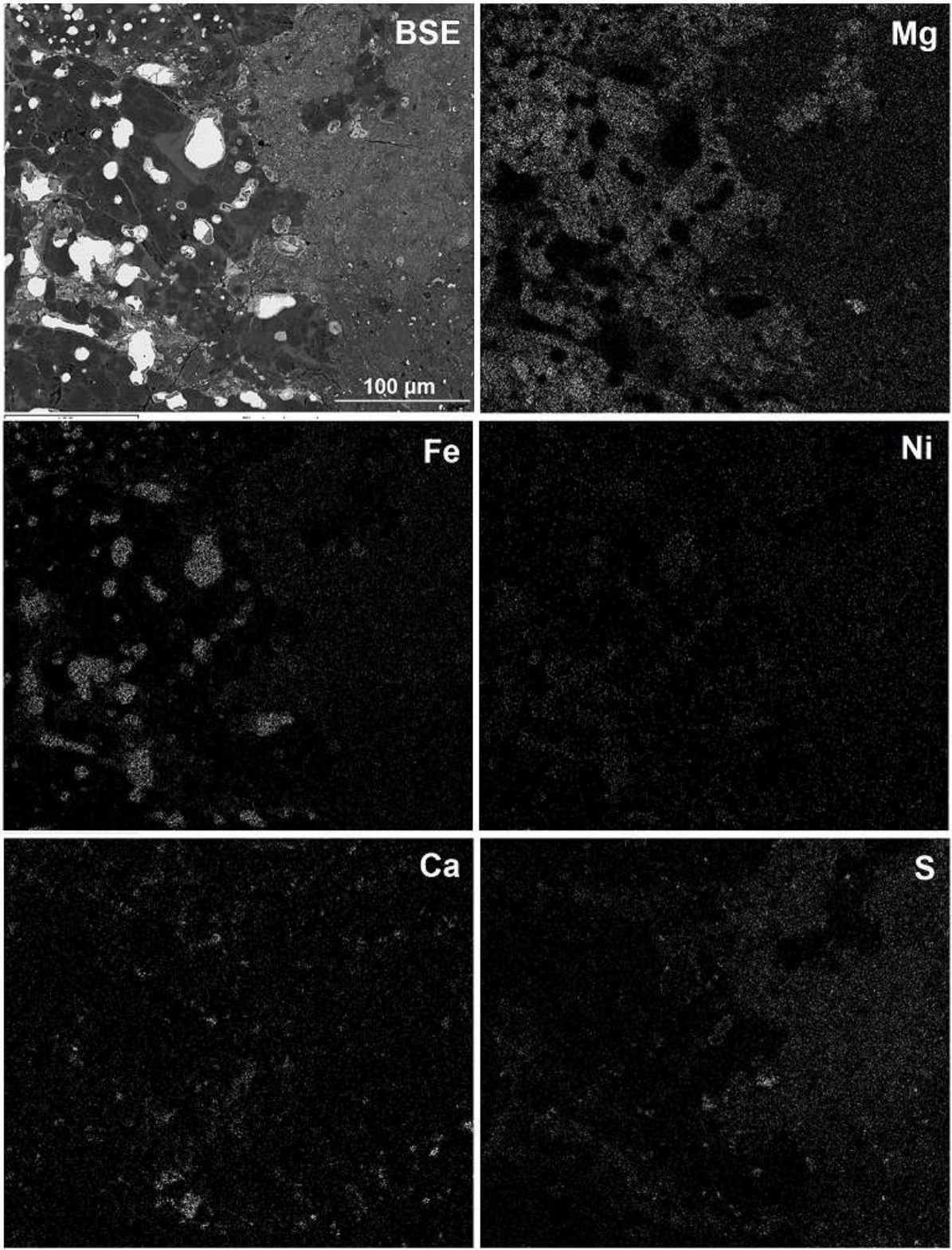

Figure 6. BSE image and Mg, Fe, Ni, Ca and S X-ray maps of the CR2 chondrite Renazzo showing evidence of aqueous corrosion of metal grains and a chondrule border.

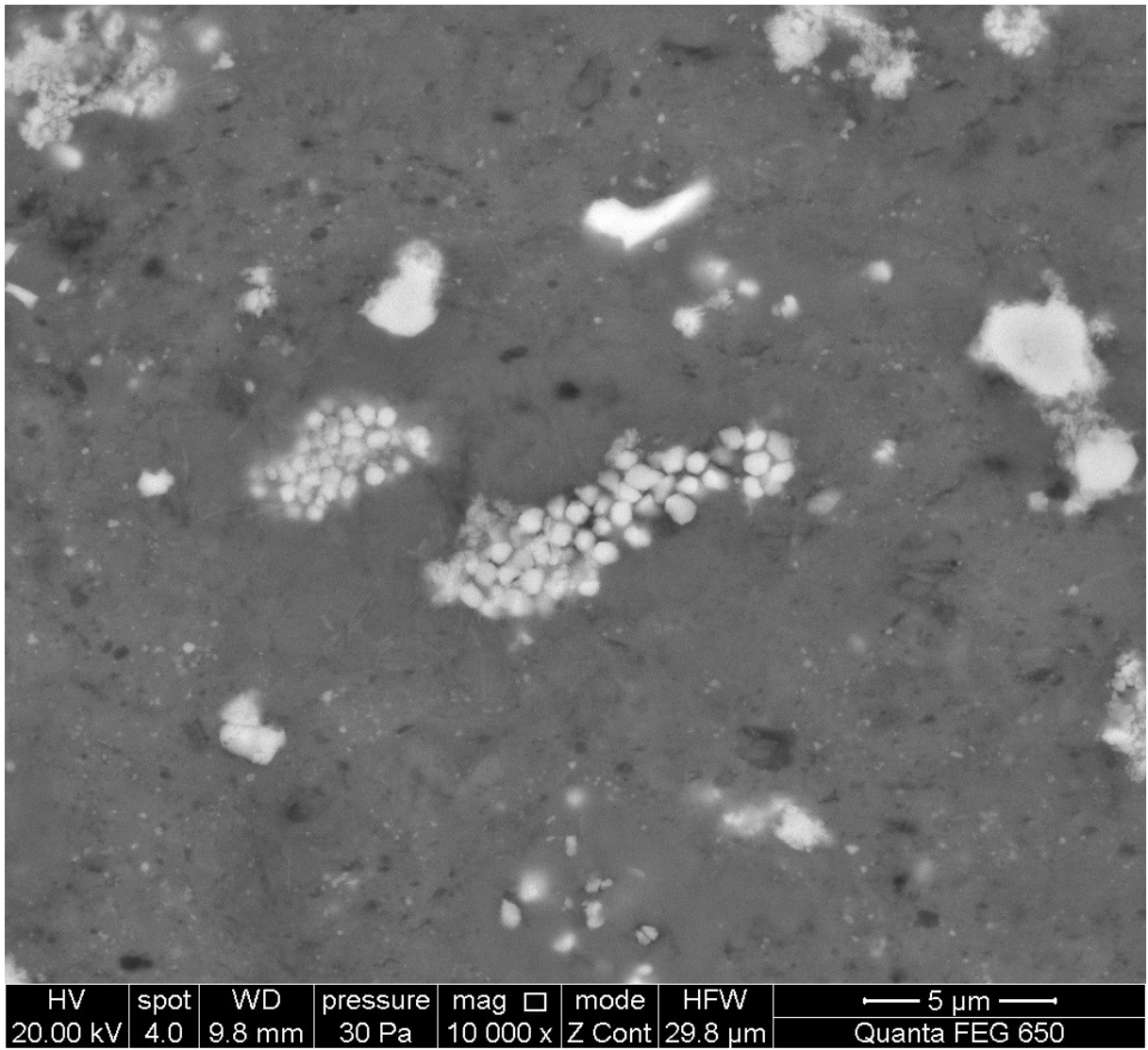

Figure 7. BSE image of the GRA 95229 CR chondrite showing pyrrhotite grains that were removed by the action of water, and the pores were filled by magnetite grains on the pore walls that belong to the meteorite matrix (for more details: Trigo-Rodríguez, 2015).

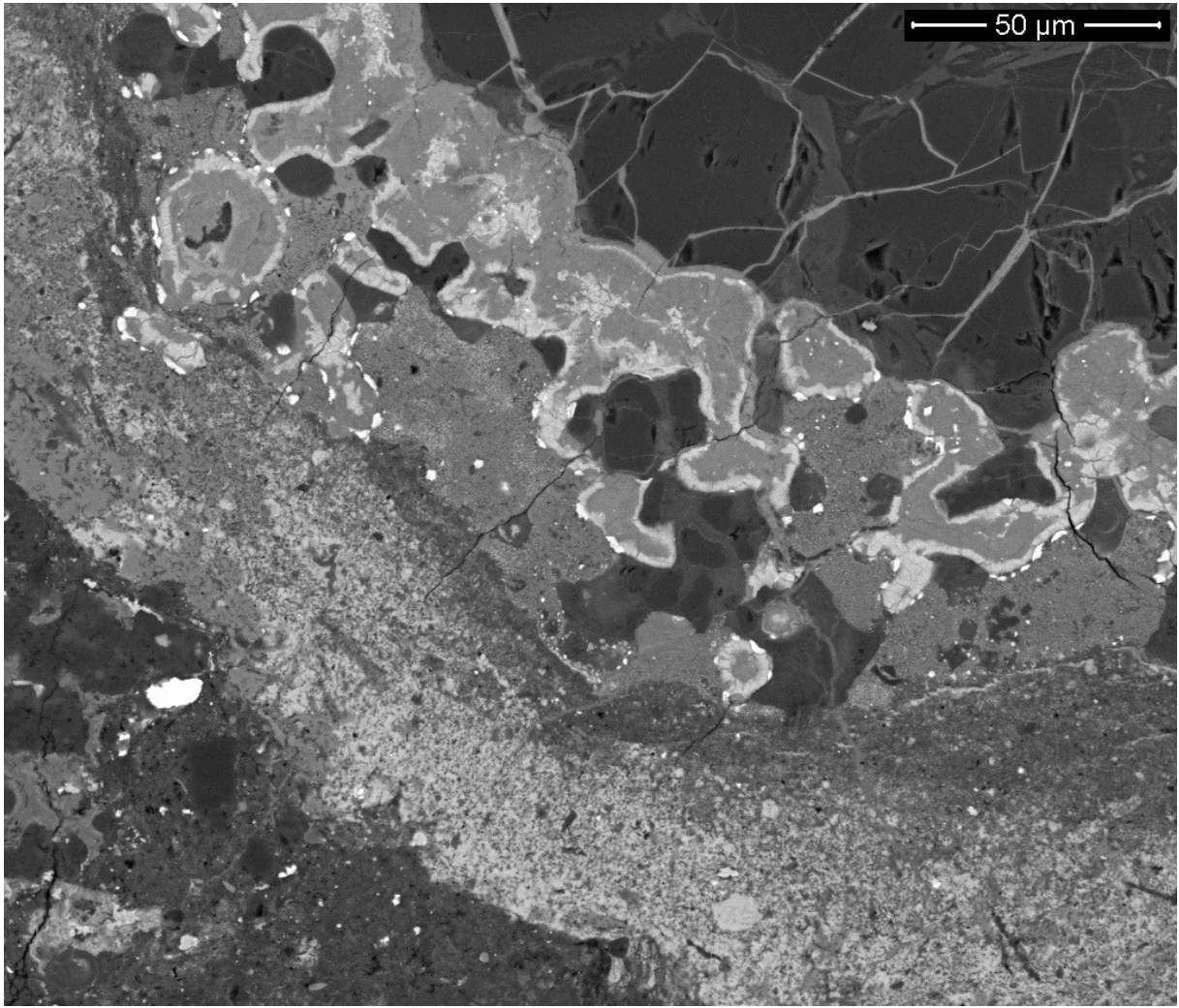

**Figure 8.** BSE image of LAP 02342 showing the fine-grained texture of a S-rich rim surrounding a chondrule (located in the upper right). It is remarkable that the outer silicate border appears "eaten" as consequence of pervasive, but localized aqueous alteration. For more details (Moyano-Cambero et al., 2016; Nittler et al., 2019)

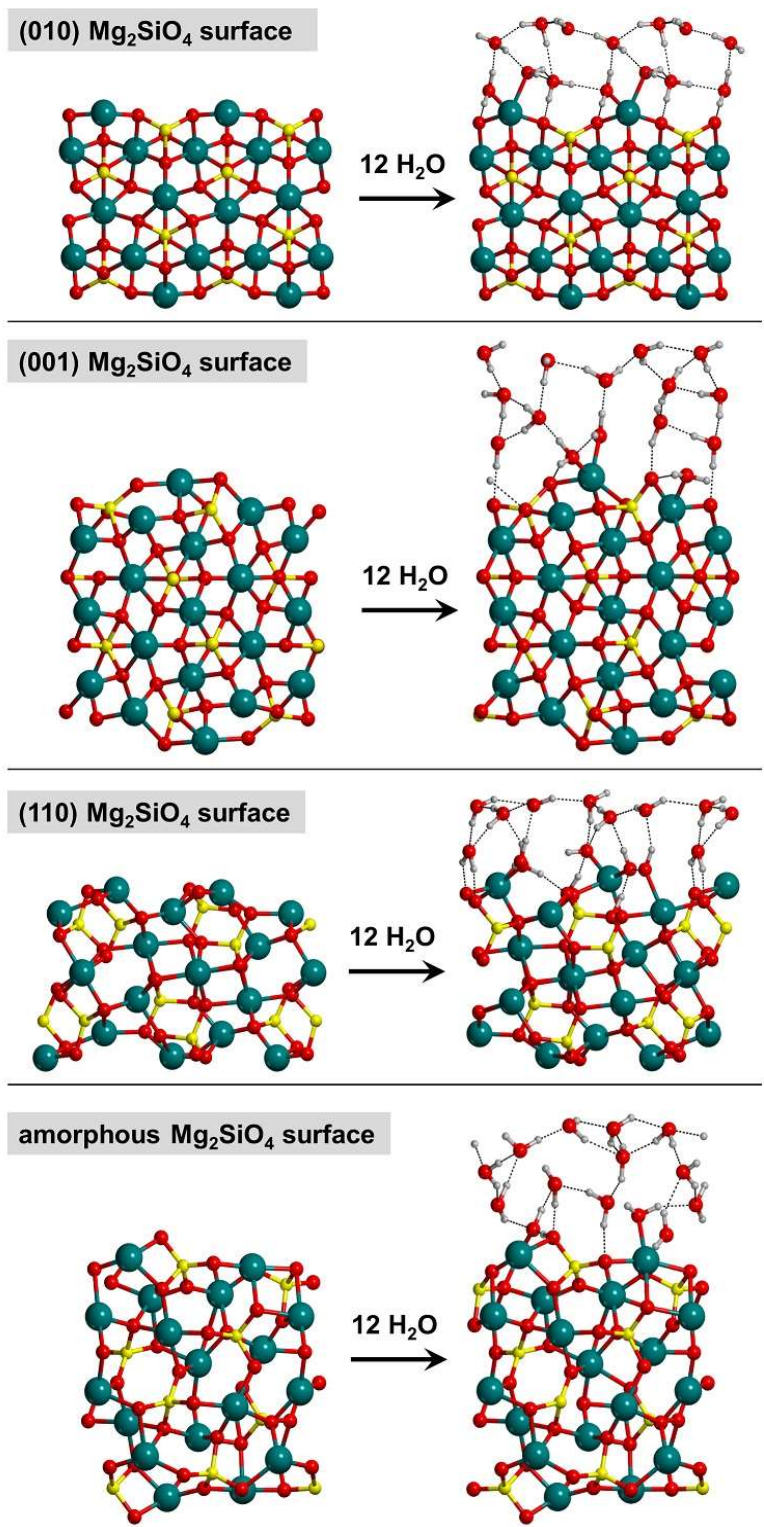

Figure 9. Clean extended $Mg_2SiO_4$ surfaces (left) and in interaction with 12 water molecules (right). For the (010) and (001) surfaces, the water molecules are molecularly adsorbed. For the (110) and amorphous surfaces, some water molecules interacting with the outermost $Mg^{2+}$ cations are dissociated forming the MgOH and SiOH surface groups (Rimola & Trigo-Rodríguez, 2017).

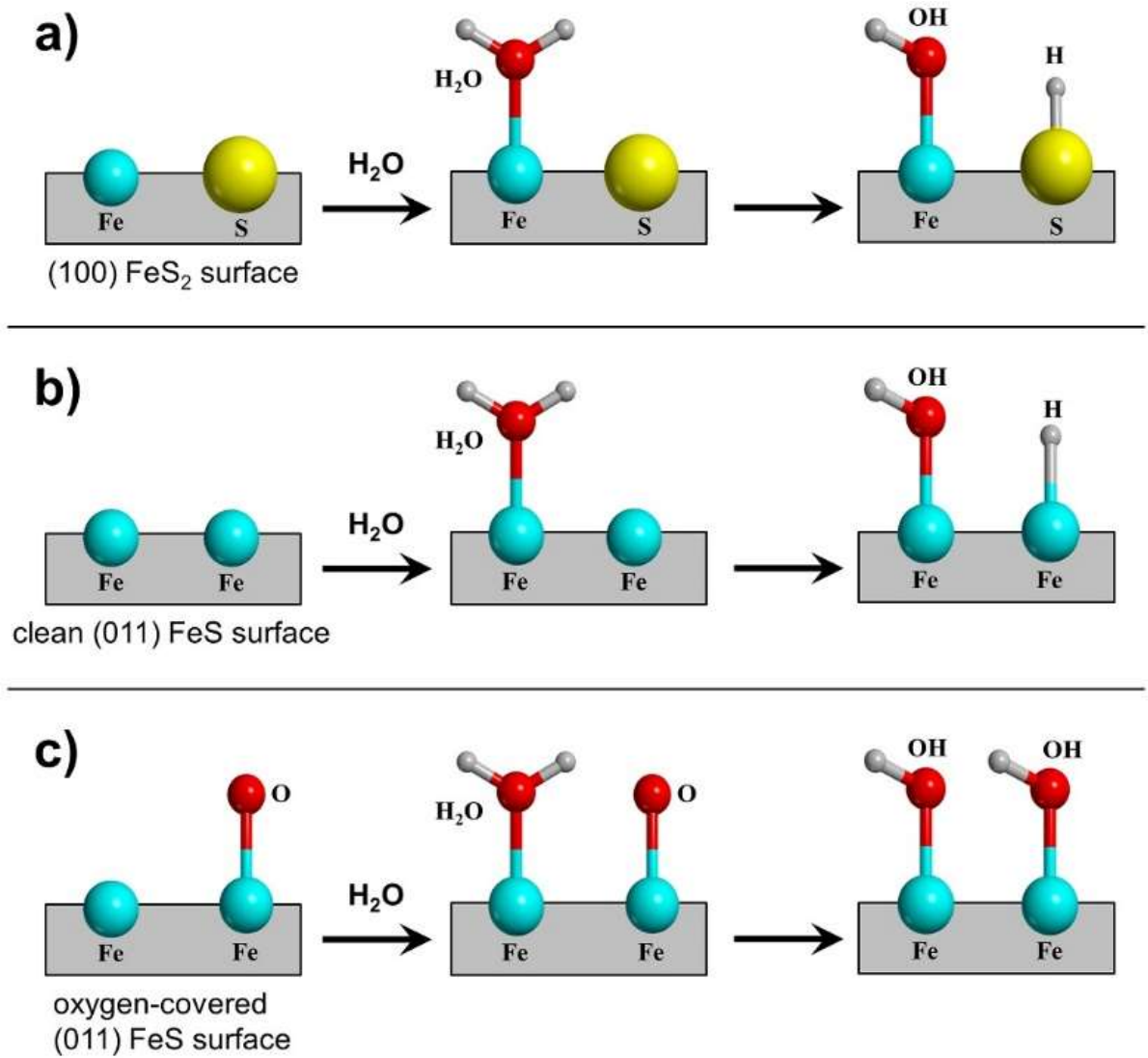

Figure 10. Schematic representation of the dissociation of water when in interaction with the (100) FeS$_2$ surface (a) (Stirling et al., 2003), and the clean (b) and the oxygen-covered (c) (001) surface of FeS Mackinawite (Dzade et al., 2016).